# Absorptive capacities and economic growth in low and middle income economies

## Muhammad Salar Khan

### Schar School of Policy and Government, George Mason University

### Email: mkhan63@gmu.edu

**Abstract:** I extend the concept of absorptive capacity, used in the analysis of firms, to a framework applicable to the national level. First, employing confirmatory factor analyses on 47 variables, I build 13 composite factors crucial to measuring six national level capacities: technological capacity, financial capacity, human capacity, infrastructural capacity, public policy capacity, and social capacity. My data cover most low- and middle-income- economies (LMICs), eligible for the World Bank's International Development Association (IDA) support between 2005 and 2019. Second, I analyze the relationship between the estimated capacity factors and economic growth while controlling for some of the incoming flows from abroad and other confounders that might influence the relationship. Lastly, I conduct K-means cluster analysis and then analyze the results alongside regression estimates to glean patterns and classifications within the LMICs. Results indicate that enhancing infrastructure crucial (ICT, energy, trade, and transport), financial (apparatus and environment), and public policy capacities is a prerequisite for attaining economic growth. Similarly, I find improving human capital with specialized skills positively impacts economic growth. Finally, by providing a ranking of which capacity is empirically more important for economic growth, I offer suggestions to governments with limited budgets to make wise investments. Likewise, my findings inform international policy and monetary bodies on how they could better channel their funding in LMICs to achieve sustainable development goals and boost shared prosperity.

## Introduction:

Economic growth across the world and, in particular, within the low- and middle-income economies (LMICs) is uneven. Some of those economies have run out ahead. Others have remained trapped in the middle, and yet many have to get out from a deep trench. Why is this the case that some economies have higher economic growth than others? This intriguing but complicated question of economic growth differential has been a subject of debate within the realm of conventional economics for decades.

Despite substantial discussion around the factors of economic growth differential, the issue is still unresolved. How these factors come to exist and how countries deploy them still needs to be investigated. Particularly, economic growth analyses need to thoroughly incorporate what Fredrick List refers to as the *capacity* of a country to absorb, exploit, and create knowledge (List 1841). However, three main conceptual and practical reasons seem to downplay the significance of knowledge absorption and capacities in economic growth studies.

First, many studies argue that knowledge absorption, exploitation, and creation are individual rather than national attributes, as illustrated in research (Heath 2001; Fagerberg and Srholec 2017). Because of this notion, these processes are not viewed as impacting economic development at the national scale. However, I contend that these are national attributes constituting national behavior. This behavior, a direct manifestation of interactions among individuals and institutions, impacts economic development. Despite their importance, these national attributes lack due prominence because it is challenging to aggregate measures of complex concepts. While some indices, such as the "competitiveness index" (WEF 2020) and the "global entrepreneurship index" (Acs, Szerb, and Autio 2017), sometimes make national media headlines and are demanded by policymakers, these measures do not often engage theory comprehensively (Im and Choi 2018; Erkkilä 2020).

The second reason why capacities and knowledge are not paid due attention in comparative analyses of economic growth is that studies diverge on what causes economic growth. For instance, studies on exogenous growth models, prioritizing factors outside of the economic system as the prime determinants of growth (Mansell 2014; Solow 1956), assign a backseat to knowledge creation and capacities adoption in the economic growth equation. Under such models, policymakers have little incentive to direct these factors. Conversely, endogenous growth and New Growth Theory studies (Romer 1994; Arrow 1962; Lucas 2004) contend that knowledge creation and capacities adoption within the system are prime determinants of economic growth. While exogenous models were dominant in the past, the recent evidence on economic growth, particularly in *Asian Tigers,* points to endogenous growth. Although these studies make important contributions to the understanding of economic growth, they do not fully appreciate and analyze the extent of national capacities and knowledge absorption that, I argue, is at the forefront of economic growth analysis in LMICs.

The last very important interrelated reason is more of a pragmatic concern. Even if researchers consider knowledge and capacities essential for creating economic value, it is extremely hard to measure and incorporate them in economic growth analyses compared to the well-known determinants of economic growth, such as *capital accumulation*.

These concerns persist secularly in the economic literature worldwide. However, the global south, especially the poorest among them, further faces underrepresentation in the economic growth literature. Part of the problem is the relevance of the existing frameworks and concepts that render them not fully applicable to social and political realities within the poorest economies. This is so because the existing frameworks and ideas are largely conceived in richer economies. Another fundamental issue for the lack of reasonable literature representation from the low-income economies stems from the lack of data. With missing data, low-income economies are excluded from the analysis. Consequently, results are not representative of the entire world economies.

Against this backdrop, I argue for a holistic approach to study growth differentials in LMICs, also earlier pleaded by Fagerberg and Srholec (2017). In concurrence with Fagerberg and Srholec (2017), my approach calls an economy (an LMIC for this analysis) a warehouse of knowledge, skills, institutions, resources, finance, and infrastructure, in other words, capacities. An essential difference in my approach is comparing a complete list of capacities *among LMICs,* primarily those supported by the World Bank International Development Association (IDA), and not with those in wealthier countries. Such capacities, I theorize, are fundamental tools to the generation of economic value in LMICs.

I define capacities as part of my proposed framework of the National Absorptive Capacity System (NACS). The capacities include business environment and finance, infrastructure (ICT, energy, and trade- and transport-related infrastructure), technology and innovation, human capital, public policy (including indicators from the World Bank's Country Policy and Institutional Assessment

CPIA clusters and other indicators of legal strength, statistical capacity, and environmental sustainability policies), and social capacity interventions (including indicators on welfare, inclusion, and equity).

For the last three decades, capacities have been extensively understood as firm-level phenomena in management and innovation literature (Cohen and Levinthal 1990; Zahra and George 2002; Müller, Buliga, and Voigt 2021; Duan, Wang, and Zhou 2020; Kale, Aknar, and Başar 2019). However, since both firms and nations are, in essence, collectives, capacities can be defined at the national level too. Therefore, in spite of visible differences between firms and nations, a researcher can still apply the ideas from firm-level literature to understand and appreciate capacities in the national context. Moreover, many similarities between these entities underpin this idea. For instance, as organized bodies, both firms and countries comprise people with varying skills and resources, interacting with each other, and creating economic value, that is distributed among the people (Fagerberg and Srholec 2017). Additionally, there is a managerial and administrative level of governance in both firms and nations that incentivize people's performance, invest in the acquisition of skills, and foster the creation and distribution of economic value. Therefore, employing firm-level concepts on a national level to advance an understanding of economic growth in LMICs offers interesting insights.

I postulate that national capacities positively and significantly impact economic growth in LMICs, after controlling for important confounders that may influence this relationship. Keeping in view data-poor environments within LMICs, I have developed a dataset for this current study (MSK dataset) from an extensive set of variables (47 variables) for a total of 82 LMICs between 2005 and 2019 (Khan 2021). In doing so, I supplement the existing data from the World Bank Group (World Development Indicators) and other sources with estimated data imputed for LMICs using cutting-edge statistical, multiple imputation, and machine learning techniques.

By employing factor analysis for dimension reduction and then conducting panel analysis to estimate the impact of capacities on economic growth, my research combines the strengths of these methodologies. Lastly, through unsupervised machine learning *Kmeans* clustering, I classify LMICs into five clusters based on capacities' strength to see megatrends for policy implications: leading, walking, creeping, crawling, and sleeping. I borrow some of the names from the neurodevelopmental phases of human life because I argue countries also follow a developmental progression. While I find that economic growth and capacities are generally higher in leading economies, followed by walking, creeping, crawling, and sleeping, the developmental progression among LMICs is not always linear.

My results indicate that improving infrastructure, finance, and public policy capacities enhances economic growth in LMICs. Similarly, I find skilled human capital boosts economic growth. On the other hand, technology and R&D spending, as enshrined in general human capital, does not affect economic growth. Lastly, I find that infrastructure capacity (particularly ICT and energy), followed by public policy capacity, and specialized human capital (including service and industry sector employment and government expenditure on education, among other things) offer the biggest bang for the buck in LMICs.

Performing analyses on the multiply imputed complete dataset, my study caters to the various problems of biasedness, face validity, missing confounders, and, most importantly, missing data, thus contributing to the body of literature. Similarly, the notions of firm-level capacities at a national level are a valuable and novel contribution to studying economic growth differential within LMICs. Alongside engaging a thorough list of capacities, another unique contribution to the literature is operationalizing the capacities such that they suit the context of LMICs.

Overall, my research develops a framework for absorptive capacity in LMICs inspired by firm-level management literature in conjunction with the national-level capacities and innovation system literature (details in Section 2). This framework elucidates how capacities impact economic growth across LMICs. The framework also operationalizes different dimensions of the concept of absorptive capacity. To test the framework and resolve data issues, I build a fresh, relatively recent, and complete dataset of 82 LMICs, employing an extended set of variables and engaging a more thorough list of capacities, their structure, and conception. Unlike previous studies, my research also accounts for important confounders (incoming flows) when analyzing the impact of absorptive capacities on economic growth. In short, proposing a novel framework, building a rigorous dataset, and employing standard quantitative approaches, my study tests more compellingly whether absorptive capacity influences economic growth outcomes after controlling for confounders, including incoming flows in LMICs.

By providing a ranking of which capacity is empirically more important for economic growth among LMICs, I advise finance and planning ministries with tight budgets to make wise investments. For instance, I suggest them to prioritize spending on infrastructure, public policy, specialized skills, and finance infrastructure and business environment capacities. Similarly, I advise international organizations like the World Bank, the United Nations, and the USAID implementing programs in LMICs, to integrate such important capacities in designing growth diagnostic and country partnership frameworks to achieve sustainable development goals and boost shared prosperity.

The rest of paper is structured as follows. In Section 2, I discuss the literature on the relationship between capacities and economic growth; in Section 3, I model this relationship in a novel framework. In Section 4, I derive hypotheses from this theory, and in Section 5, I discuss the data. Section 6 addresses capacities measurement, in particular, the way in which I have employed factor analysis on a comprehensive set of variables to derive composite factors to measure national capacities. The following two sections explore the longitudinal impact of the national capacities on economic growth and classify LMICs according to the capacities prior to conducting sensitivity analyses of the results. Lastly, Section 9 presents the conclusions and implications of this study.

## 2. Capacities and Economic Growth Literature:

My research focuses on the role of capacities in economic growth in LMICs eligible for the World Bank's IDA support. Researchers in different fields have studied economic growth differential across countries. Historically, there is a range of perspectives from Solow's (1956) view of differences in the amount of accumulated capital per worker to Gerschenkron's (1962) idea of technological differences, and then later on the same notion by the various advocates of "new growth theory" (Romer 1994; Lucas 2004) ascribing economic development differences across countries to variation in the degree of technology adoption and human capital accumulation. Empirical works on the industrialization processes in Asian and Latin American countries undertaken during the 1970s and 1980s demonstrate an active government role in developing "capabilities" (here termed as capacities) required to catch up (Kim 1980; Fransman 1982; Dahlman et al. 1987; Lall 1992). In that period, the concepts of "technological capability" (Kim 1980, 1987) and "social capability" (Abramovitz 1986) emerged to explain development. Since these concepts provide foundations to many works linking capacities and economic development, it is important to illustrate them briefly.

Kim (1997) defines technological capability as "the ability to make effective use of technological knowledge in efforts to assimilate, use, adapt and change existing technologies." Further, he asserts that technological capability has three aspects: innovation capacity, production capacity, and investment capacity. Hence, Kim's concept includes both planned R&D—which arguably is a small activity in many LMICs—and capacities necessary to exploit technology on a commercial scale. On the other hand, social capability encompasses societal or collective capacities regarding what organizations can do and how this is aided (or impeded) by broader social and cultural factors. Among the aspects of social capability that Abramovitz (1986) highlights include technical skills, experience in organizing and managing large-scale enterprises, working financial institutions, markets mobilizing capital, honesty and trust, and governments' stability and ability to make and enforce rules and support economic growth. While these concepts are useful characterization of important ideas, they typically lack a rigorous operationalization, which is key to testing theories of their impact on economic growth and relative importance.

Around the same time, a more interactive approach in the form of a "national innovation system" (NIS) surfaced, focusing on systems, activities, institutions, and their interactions as the driving force of growth and development (Lundvall 1992; Nelson 1993; Edquist 1997; 2006). The NIS literature also utilized the concepts of technological and social capacities (Castellaci and Natera 2011). Most of the initial theoretical and empirical work on NIS focused mainly on prosperous economies (Nelson 1993; Edquist 2001); however, later theoretical NIS research became more inclusive by emphasizing "diffusion," "imitation," and "learning" processes in developing economies (Viotti 2002; Lundvall et al. 2009; Casadella and Uzunidis 2017). This literature termed developing LMICs as "national economic learning" entities and "imitation centers (Viotti 2002; Lundvall et al. 2009; Fagerberg and Verspagen 2002; Casadella and Uzunidis 2017). Despite this emphasis on diffusion and learning, this NIS literature, in general, has not fully explained absorption and learning processes and how they might happen in developing economies. Since the concept of NIS was primarily conceived in prosperous economies, NIS literature, by and large, does not capture the social and political realities within the developing LMICs.

Around the time NIS emerged, Cohen and Levinthal (1990) developed the idea of "absorptive capacity" to explain how learning is consolidated at the firm-level and how it impacts a firm's growth. They define absorptive capacity as "the ability of a firm to recognize the value of new, external information, assimilate it, and apply it to commercial ends." Other researchers termed this concept a multifaceted and *complex construct* (He et al. 2015; Minbaeva et al. 2014; Fagerberg and Srholec 2008). Over the past three decades, it gained considerable traction in many fields, including strategic management, international business, and organizational sciences (for example, see bibliometric analysis of Absorptive capacity by Apriliyanti and Alon 2017; César and Forés 2010; (Kale, Aknar, and Başar 2019; Müller, Buliga, and Voigt 2021). Most of these works included *acquisition, assimilation, transformation,* and *exploitation* as dimensions of absorptive capacity at the firm level. Researchers investigating firm-level processes have envisioned and operationalized these dimensions in various interesting ways.

Some of the firm-level studies proxied absorptive capacity in terms of a firm's R&D investment; such studies argue that through its R&D activities, a firm develops collective knowledge about specific areas of markets, science, and technology (Aldieri, Sena, and Vinci 2018; Omidvar, Edler, and Malik 2017; Brinkerink 2018). Subsequently, the firm then employs the knowledge gained in the design and development of its products and services, eventually increasing its economic value (Brinkerink 2018). However, the R&D indicators on firm-level are only great in richer economies; they are weak indicators of firms' performances in LMICs because those firms rarely have any R&D budgets.

In the early 2000s, Narula (2004) and Criscuolo and Narula (Criscuolo and Narula 2008) extended the firm-level concept to a national level. They developed a theoretical framework for aggregating national absorptive capacities upwards from the firm level. Aggregating individual firms' absorptive capacities to understand national-level processes seems a workable idea. However simple as it sounds, it is inappropriate to aggregate individual firms' absorptive capacities for two reasons. Firstly, national level processes or national absorptive capacities are more than the sum of the absorptive capacities of domestic firms or industries. Various multiplicative effects from interacting with a globalized world, perhaps insignificant at a firm level, become very significant at the national level. Just as firms operate within systems, countries are not isolated from outside knowledge. The external technological environment, foreign knowledge and aid, and the stock of knowledge of firms of other countries also influence national absorptive capacity processes; thus, these variables need to be considered. Secondly, aggregation from the firm level not only may underestimate (because of multiplicative effects), but simply not capture national-level processes.

For instance, the firm-level aggregation completely misses the national regulatory environment, government's capacity, national fiscal and financial management, legal system, infrastructure, and business enabling environment, among other things. Since these things are not even firm characteristics, no such aggregation would capture them.

Other recent empirical studies applied the idea in a national setting (Fagerberg and Srholec 2008 and 2017). These works have also used different capacities earlier applied in the NIS literature (such as technological and social capacities) as proxies or measures for absorptive capacity. Such studies are not very comprehensive particularly when applied to study absorptive capacities in LMICs, because absorptive capacity includes many other well-specified capacities (such as finance, business environment, infrastructure, manufacturing, and service sector employment, among other things) that may impact economic growth in LMICs.

Some national-level studies (Fagerberg and Srholec 2008 and 2017) have extended the set of capacities. However, they still have been using R&D indicators (like firm-level studies), and other variables that I argue are not best in capturing the absorptive capacity processes in LMICs. For example, after conducting factor analysis on 25 indicators and 115 countries from the 1992-2004 period, Fagerberg and Srholec (2008) identified four different types of capacities: the development of the innovation system, the quality of governance, the type of political system, and the openness of economy. The authors concluded that innovation systems and governance were particularly crucial for economic development. Similarly, they conducted another study on 11 indicators covering 114 countries worldwide on different levels of development for the period 1995-2013 (Fagerberg and Srholec 2017). After factor analysis, they grouped the indicators into three capacities: "technology," "education," and "governance." They found technology and governance as significant for economic development.

While these studies provide a starting point, their indicators do not have strong measurement validity, particularly when the countries under study are the poor LMICs. For example, using R&D investment and journal articles for technology and innovation may not fully capture innovation as innovation does not entirely manifest itself in R&D investment or journal articles in poor economies. R&D proxies are also not suitable because R&D expenditures and allocations are seldom paid attention to in poor economies. Moreover, these economies may be allocating just sufficient R&D, but they do not know how to utilize R&D for beneficial activities because of the lack of an enabling environment. Similarly, S&T articles do not capture absorptive capacity in poor economies because most of these articles do not translate in any significant value for many reasons. A prime reason is that producers and innovators in these economies hardly utilize the results of scientific research directly to produce economic value. Likewise, Law and Order and (lack of) corruption produce a limited measure of governance that ignores important governance characteristics; there are better ways of measuring it.

Secondly, these studies did not consider vital indicators that could correlate with treatment capacities and the outcome variable. For instance, they omitted important confounders such as incoming flows from abroad (foreign aid, technical and cooperation grants, among others), committing omitted variable bias. In the presence of such a bias, some capacities seem to matter more or less than they do.[1]

Thirdly, while they have included countries with varying levels of development, the estimates are not representative of all the nations because their analyses excluded many developing low-income economies due to missing data. Fourthly, they do not consider important financial, bureaucratic, economic environment and information and communications technology (ICT) infrastructure and institutional factors and policies, which can be crucial for a developing country's absorptive capacity and subsequent development.

All these firm-based and NIS approaches, at best, only capture part of reality, the country's absorptive capacities. A more comprehensive framework is needed to measure and capture accurately and comprehensively dimensions of absorptive capacity at the national level in developing LMICs.

Another big shortcoming of the current national-level empirical studies lies in methodological challenges. Empirical studies of capacities and development have used mainly two methodologies:

---

[1] For instance, Fagerberg and Srholec (2017) only include three dimensions of Governance: government effectiveness, (lack of) corruption, and law and order. The estimated coefficient size for the impact of governance on GDP per capita is 29%. This seems most likely an overestimation because governance includes many other important indicators on institutions and public sector management, which they do not have in their analysis. Similarly, they include tertiary, secondary, and primary attainment as measures for Education capacity. The estimated coefficient size on overall education is 0, which is likely an underestimation because education capacities include many other indicators, which they overlook. Perhaps, because of the exclusion of many important variables, their models also lack overall explanatory power: the best model only explain about 43% of variation (R-squared value = 0.43).

panel regression analyses (Teixeira and Queirós 2016) and composite indicator analyses (Fagerberg and Srholec 2017). While these analyses can handle many variables, countries, and periods, data availability imposes severe limitations.

Panel regression analyses account for a few key variables that supposedly measure countries' differences in their different capacities. Subsequently, these studies examine the empirical relationship between these variables and comparative national differences in GDP per capita growth (Castellacci, 2004, 2008 and 2011; Teixeira and Queirós 2016; Ali, Egbetokun, and Memon 2018). While powerful as they consider the dynamic nature of capacities, such panel studies particularly ignore many low-income economies because longitudinal data for many variables are missing in these countries. These analyses drop off the countries for which there are missing data for variables through listwise deletion. As a result, the coefficients of interest obtained through panel analyses do not provide information about the poorest economies. The estimates obtained through such studies may exhibit an upward bias by overestimating the effect of *capacities* on *economic growth*.

On the other hand, composite indicator analyses build aggregate indicators and conduct descriptive analyses. Such studies use many variables, denoting various dimensions of technological and social capacities. The variables are then systematically combined into a single composite indicator through factor and cluster statistical tools (Fagerberg and Srholec 2008; 2015; 2017). The composite indicator is a country's comparative standing against other countries. As opposed to panel analyses, the composite analyses consider many countries, including some low-income economies. But since low-income countries have limited data available, such studies are usually static (one-year study), ignoring system-level evolution in the countries analyzed. Additionally, not all low-income economies have data on all the variables of interest available for one particular year. Therefore, even composite analyses cannot possibly include all low-income countries.

Keeping in mind the issues of conceptual relevance and data and methodology challenges, my research establishes a renewed relationship between conceptual and empirical work. For conceptual understanding, Fagerberg and others consulted foundational literature on "technological" and "social" capacities instead of employing the framework of absorptive capacity from the strategic management (and business) literature. It seems like their understanding was driven primarily by choice of their methodological approach and data availability. My research, on the contrary, develops a framework for absorptive capacity in LMICs informed by strategic management literature in conjunction with the national-level capacities and NIS literature. This framework illuminates how capacities impact economic growth across LMICs. The framework also measures various elements of the concept of absorptive capacity. In order to test the framework and settle data issues, I construct a fresh, relatively recent, and full dataset of 82 LMICs, utilizing an expanded set of variables and employing a more thorough list of capacities, their structure, and conception. In contrast to previous studies, my research also considers key controls (incoming flows) when analyzing the impact of absorptive capacities on economic growth. In summary, offering a novel framework, building a rigorous dataset, and engaging established quantitative approaches and tools, my study tests more thoroughly whether absorptive capacity influences economic growth outcomes after controlling for controls, including incoming flows in LMICs.

## 3. The National Absorptive Capacity System (NACS) Framework:

A rich analytical framework is needed to capture capacities and their impact on economic growth, befitting the data deficient environments of the LMICs. I call the proposed framework, *National Absorptive Capacity System (NACS).* The framework for National Absorptive Capacity System (NACS) situates a developing nation as an "economic learning" entity, constantly absorbing, exploiting, and using knowledge, skills, and learning and converting the gains into economic value proportionate to the strength of its "local" capacities. I develop NACS based on the firm-level concept of "absorptive capacity" found in the strategic management literature (Cohen and Levinthal 1990; Zahra and George 2002). From the firm-level literature, I consider a nation an analogous entity where individuals and institutions interact, learn, create, and distribute economic value according to some set rules. Further, the National Innovation System (NIS) literature also inspires this NACS framework. The NIS literature, particularly its later literature on emerging economies, informs that a developing nation, as an active learning entity, absorbs and creates knowledge and improvises on the existing knowledge (Casadella and Uzunidis 2017; Juma et al. 2001).

While they jointly provide foundation and legitimacy to the framework I propose, the firm-level absorptive capacity and traditional NIS concepts are by themselves insufficient and inappropriate to fully and accurately capture absorption processes and their subsequent impact on economic growth in developing LMICs, as illustrated in the literature review section. The proposed NACS framework thoroughly envisions capacities and absorption processes in LMICs and rigorously operationalizes the capacities, employing concepts from firm-level and national-level NIS literature. The following subsection describes the development of the NACS framework.

### 3.1 *Developing the Framework for Absorptive capacity: From Firm to Nation*

My NACS framework consists of three main elements. The first central element is "absorptive capacities," which this research examines. The second element is "outcome processes," which are hypothesized to be impacted by capacities. Finally, the third element is "control inputs," which may influence the relationship between the first and second elements. Figure 1 below depicts this framework:

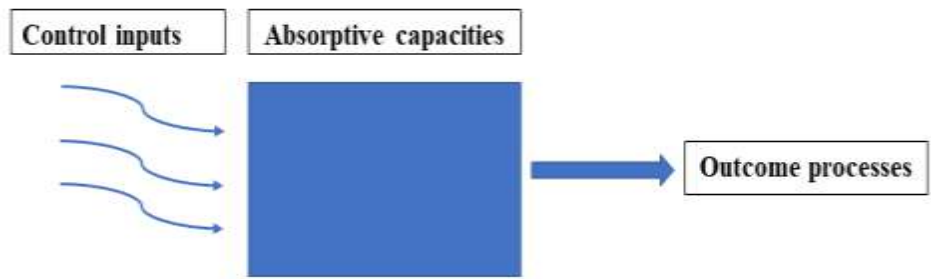

**Figure 1: National Absorptive Capacity System—a framework showing how absorptive capacities influence outcome processes while controlling for confounders or controls (control inputs).**

More formally, I illustrate this figure in the form of the following equation (or model):

$$Log(Y) = \Sigma\, \alpha_n C_n \,+\, \Sigma\, \beta_n Z_n \,+\, \varepsilon \qquad\qquad (1)$$

In equation (1), $Y$ shows outcome processes, $C_n$ indicates all absorptive capacities, $Z_n$ shows confounders or control variables (including incoming flows), and the symbol $\varepsilon$ shows the error term. Finally, vector coefficients $\alpha$ and $\beta$ measure the impact of capacities and confounders on outcome processes, respectively.

To define these elements of national absorptive capacity, I employ firm-level dimensions of absorptive capacity: knowledge and skills *acquisition*, *assimilation*, *transformation*, and *exploitation*. These dimensions translate into related yet very different things when applied to the national level.

Let us define the firm-level dimensions and respective national level elements derived from these dimensions:

### I.        *Control inputs—Acquisition and Assimilation*

On a firm level, the *acquisition* is a company's ability to capture external knowledge based on its efforts (Cohen and Levinthal, 1991). Similarly, *assimilation* is the absorption (internalization and diffusion) of *acquired* external knowledge (Zahra and George 2002).

On a national level, I term both of them as control inputs. I see them as the ability of a nation to make a deliberate effort in capturing and assimilating external knowledge (learning, training, technology, and skills), practices, and resources. While it is hard to measure the extent or magnitude of such inputs, the size of *incoming* flows informs about their strength. Similarly, other controls, such as country geographical status (landlocked vs. coastal), natural resources, and population density, influence the rate of acquisition and assimilation. Thus, this formal model includes incoming flows and other controls to indicate acquisition and assimilation on a national level. Relevant variables to measure the two are population size, geography, resources, brain flow, linkages, officials flow, and information flow.

The framework here includes control inputs such as population size, capital formation, technological cooperation grants that LMICs receive from developed countries and donors, international tourist arrivals, merchandize import from the high-income economies, and net Official Development Assistance (ODA) received from abroad.

### II. *Absorptive capacities—Transformation*

Transformation on a firm-level refers to the combination and recombination of old and new knowledge in the pursuit of adding value (Vasconcelos et al. 2019; Müller, Buliga, and Voigt 2021). Transformation is proportional to a firm's competencies and resources. Extending the concept of transformation to a national level implies internal capacities (and their adoption), including infrastructure, in realizing economic outcomes. Such capacities help construct new routines, new products, new processes once the new knowledge is *assimilated* and spread in a country. The NIS literature provides valuable insights here. This literature considers technology, governance, human capital, and infrastructure, among others, as prime capacities. My framework in this paper includes six capacities drawn from the literature: 1) Technological capacity, 2) Financial capacity, 3) Human capacity, 4) Infrastructural capacity, 5) Public Policy capacity, and 6) Social capacity. Figure 2 refers to these capacities while also acknowledging the incoming flows held constant in this framework.

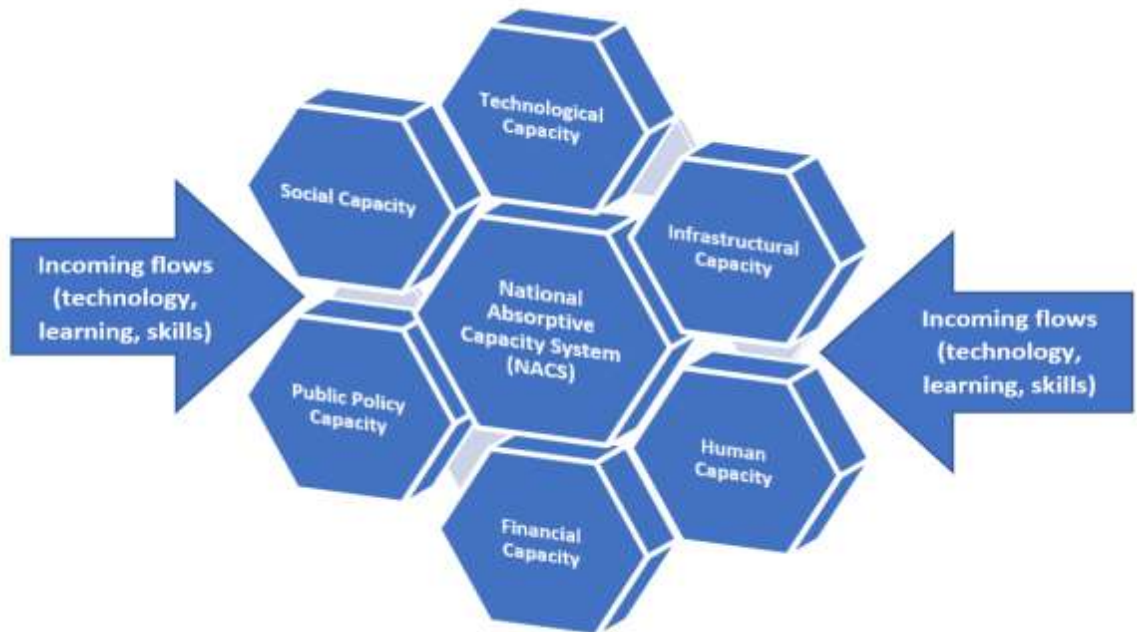

**Figure 2: National Absorptive Capacity System (NACS) and its capacities (Khan 2021). These six capacities constitute the bulk within the framework for NACS. Incoming flows, held constant within NACS, are also shown.**

### III.    *Outcome processes—Exploitation*

After a firm *acquires, assimilates, and transforms* knowledge, the firm moves towards knowledge application (Solís-Molina, Hernández-Espallardo, and Rodríguez-Orejuela 2018; Seo, Chae, and Lee 2015). Exploitation or knowledge application on a firm-level entails creating new products or services by using competencies or improving competencies (Cohen and Levinthal 1990). Gebauer and Colleagues (2012) measure it by employing the commercial application of the acquired knowledge. Similarly, César and Colleagues (2010) consider exploitation as activities related to product and process changes and improvements. On a national level, this step indicates achieving a country's economic outcomes by engaging the capacities. Economists usually measure economic outcomes in terms of economic growth. Thus, exploitation can be indicated by many of the outcome processes, including economic growth (GDP growth), per capita GDP, value-added industry growth, product, process, marketing, or organizational innovations. I consider the GDP per capita (and GDP per capita growth) of countries as an outcome for pragmatic concerns.

To sum it up, the NACS framework illustrated here (and shown in Figure 1) combines the firm-level *acquisition* and *assimilation* into confounders or control inputs whereas, firm-level *exploitation* signifies national outcome processes. Finally, the firm-level *transformation* is a BlackBox on a national level that includes national capacities influencing outcome processes.

Based on this framework, I test the following hypotheses in the context of LMICs.

## 4. Hypothesis:

The literature so far suggests that various factors (dimensions) within capacities positively impact economic growth. For instance, science and technology indicators (within technological and human capital capacities) have proven to positively impact economic growth (Çalışkan 2015; Wu, Zhao, and Wu 2019; Pinto and Teixeira 2020; Baneliene and Melnikas 2020; Bhalla and Fluitman 1985; Laverde-Rojas and Correa 2019). Similarly, in developing economies, functioning financial markets catering to citizens and businesses and financial inclusion also improve economic growth (Durusu-Ciftci, Ispir, and Yetkiner 2017; Asteriou and Spanos 2019; Ibrahim and Alagidede 2018; Kim, Yu, and Hassan 2018).

Furthermore, developing countries need robust ICT infrastructure, energy access, and transport-related infrastructure to have strong economic growth (Bahrini and Qaffas 2019; Munim and Schramm 2018; Saidi, Shahbaz, and Akhtar 2018; Mohmand, Wang, and Saeed 2017).

In addition, social capacity (including redistribution and income equality) across the world has been found to improve economic growth (Kennedy et al. 2017; Berg et al. 2018). Lastly, public policies in terms of fiscal management, financial management, bureaucratic management, strong institutions have been affecting economic growth across the world (Hussain et al. 2021; Urbano, Aparicio, and Audretsch 2019; Williams 2019; Acemoglu and Robinson 2019; Alexiou, Vogiazas, and Solovev 2020; Acemoglu, Johnson, and Robinson 2005; Asghar, Qureshi, and Nadeem 2020). Without an effective public policy, a country might not experience the fruition of other capacities.

However, such capacities are not the only factors influencing the economic growth of a developing country. In this globalized world where a nation does not operate in isolation, incoming flows such as technical cooperation grants, foreign aid, and merchandise import from high-income economies and other controls such as a country's population density and capital stock also may influence economic value (Galiani et al. 2017; Asongu and Ezeaku 2020.; Kugler 2006; Uneze 2013; Bal, Dash, and Subhasish 2016). For instance, one such inflow of technical cooperation grants[2] was found to positively (and jointly with loan aids) influence economic growth in Sub-Saharan Africa (Asongu and Ezeaku, n.d.).

Based on this literature and in light of the NACS framework, my hypothesis is:

**H:** *Capacities (defined within the transformation stage of NACS) positively influence economic growth in LMICs after controlling for confounders (including incoming flows).*

While economic growth is captured by the natural log of GDP per capita (outcome variable), the capacities (used as independent variables) must be measured and operationalized (more on capacities and conception in Section 6). Before explaining the process of measuring capacities, let us briefly describe the data at hand.

## 5. Data Description:

Since capacities change over time and are multidimensional, I compare capacities and their impact on economic growth within LMICs, using a panel dataset with 82 LMICs between 2005 and 2019. Subject to a researcher's understanding and judgment, multidimensional capacities can include many variables that require dimension reduction in an organized fashion to produce intelligible results while retaining complexity. To conduct such analyses, it is necessary to have access to as many relevant variables as possible to capture all the facets of a capacity for a reasonably large set of countries and a sufficiently long time period.

Data on all the relevant variables for long timespans do not exist for many countries, particularly the LMICs, which are the prime foci of this paper. I argue such countries exhibit a data pattern, termed as *missing at random (MAR)*. The pattern, by definition, implies that the *missingness* pattern in data is conditional on *observed* variables (Afghari et al., 2019). In other words, missingness can be predicted by the observed data. LMICs can have missing data for many reasons, ranging from poor data infrastructures and meager resources to frequent natural disasters and severe civil conflicts. However, despite missingness in many variables of significance, such countries offer rich information on poverty indicators, economic development, literacy rates, and demographics. I argue that this rich corpus of data can be employed to explain and predict the missingness pattern for data on other variables, thus justifying the MAR assumption. In other words, missing values of variables in those countries are conditional on the data I observe.

---

[2] Technical cooperation grants comprise of: (i) free-standing technical cooperation grants projected for financing the transfer of technical as well as managerial skills or of technology to build-up general national capacity without reference to any explicit investment projects; and (ii) investment related technical cooperation grants, which are made available to strengthen the capacity to carry out specific investment projects ("Technical Cooperation Grants (BoP, Current US$) | Data Catalog" n.d.).

Relying on this MAR pattern of data in the LMICs, I use a *multiple imputed* MSK Panel dataset (Khan 2021) obtained after applying Rubin's Multiple Imputation by Chained Equations (MICE), specifically MICE predictive mean matching (Akmam et al., 2019). While respecting the structure of multivariate continuous panel data at the country level, this technique generated a dataset with no missing values.

For the data structure, Castellaci and Natera (2011) inspired my work (CANA hereon). They estimated dataset for 134 countries between 1980 and 2008 using a Multiple Imputation (MI) algorithm developed by Honaker and King (2010). I also applied Rubin's novel M1 techniques (Rubin 1996 and 1987) to estimate the MSK panel dataset for this study.

The MSK dataset is different from the CANA dataset. First, as opposed to the CANA dataset, the MSK dataset focuses on relatively more data-deficient and economically poor 82 LMICs. Among them, 74 countries are the most impoverished countries (low GNI per capita) eligible for the World Bank's IDA support.[3] Another eight countries included recently graduated from IDA-eligibility.[4] Second, the MSK dataset includes an extended set of other relevant variables to measure capacities. For example, MSK consists of 47 variables for all economies in the dataset. In contrast, CANA

---

[3] Eligibility for IDA support depends mainly on a country's relative poverty. Relative poverty is defined as GNI per capita below an established threshold, and it is updated annually ($1,185 in the fiscal year 2021). IDA also supports some countries, including several small island economies, that are above the operational cutoff but lack the creditworthiness needed to borrow from the International Bank for Reconstruction and Development (IBRD). Some countries, such as Nigeria and Pakistan, are IDA-eligible based on per capita income levels and are also creditworthy for some IBRD borrowing. They are termed as "blend" countries. http://ida.worldbank.org/about/borrowing-countries

[4] IDA eligibility requires a threshold GNI per capita, countries graduate and reinter (reverse graduate in the list)

consists of 34 variables for all economies and another seven variables for a restricted set of countries within the dataset.

Third, the timeframe for this study spans around fifteen years (between 2015 and 2019), not only because it is a decent period for panel analysis but also for pragmatic concerns regarding data availability, particularly on my public and social policy capacity variables. The World Bank Group's country offices started collecting these variables in the IDA-eligible countries from 2005 onwards ("Country Policy and Institutional Assessment" 2014).

The 47 variables included in the dataset were collected from publicly available databases (see Appendix Table A.1 for variables' definitions and their sources). Table A.1 also has details of a set of control variables and the outcome variable. A summary description of all the variables is available in Appendix Table A.2. The variables included in the dataset are crucial for measuring six capacities alongside incoming factors (Figure 2) included in the National Absorptive Capacity System (NACS).

The variables included are a mix of variables, primarily continuous variables and indices, measured in many ways. For instance, the outcome variable of GDP per capita is a continuous variable, constant in 2010 US dollars. Public Policy and social capacity variables are generally clustered averages and composite indicators, with low values or scores indicating lower magnitude or strength of the variables (e.g., economic management cluster ranging from 1 to 5.5, with 1 showing the low score and 5.5 meaning high score). Some variables in the financial capacity are continuous, measured in days (e.g., days to enforce a contract). Additionally, some continuous variables in technological capacity are measured per 1 million people (e.g., number of researchers or technicians in R&D). Lastly, some continuous variables in infrastructure capacity are measured per 100 people (e.g., telephone and mobile phones subscriptions).

The next section shows how I use these variables to construct capacities factors.

## 6.  Measuring National Capacities in the Framework for NACS:

I measure national capacities by constructing composite factors. To build the factors, I employ a set of relevant variables. The variables capture phenomena of interest—latent factors to be discovered in this case, and by extension, the proposed capacities in the NACS framework. I use factor analysis to generate a small number of factors from a set of many variables (Stephenson 1935; Yong and Pearce 2013). The core assumption is that variables relating to the same dimension of reality strongly correlate with each other (Bandalos and Finney 2018). Most variables are correlated from a higher to a moderate level in the current dataset, suggesting a piece of crude diagnostic evidence for factor analysis.[5] Readers interested in details about factor analysis can consult practical resources (Bandalos and Finney 2018; Yong and Pearce 2013; Goretzko, Pham, and Bühner 2019), but overall, bear in mind that correlation matrices guide this analysis.

Generally, factor analysis is conducted in one of two ways: 1) exploratory factor analysis and 2) confirmatory factor analysis. While exploratory factor analysis is an unsupervised analysis and does not require a priori input from theory or hypotheses, confirmatory factor analysis is a supervised method that incorporates prior information from theory (Bandalos and Finney 2018). Since the extant literature informs what variables might constitute different capacities, I first conduct confirmatory factor analysis. Based on my understanding of theory and literature, I assign the variables under each capacity and then employ factor analysis to reduce the variables in latent factors. For robustness, I also conduct exploratory factor analysis without any assignment of variables. Both analyses, by and large, produce similar results. I execute these analyses using *STATA* (version SE 15.1).

To explain confirmatory factor analysis more, in the first step, the literature informs to assign the variables in the MSK dataset to one of the appropriate six capacities. After assigning all variables to the capacities, I program the software to perform a total of six confirmatory factor analyses, one for each capacity, using the principal-components factors with the orthogonal varimax rotation (Chavent, Kuentz-Simonet, and Saracco 2012). The analyses return factor loadings and factors. Factor loadings indicate how each variable is related to each latent factor (see Appendix Tables B.1-B.6 for factor loadings of the six capacities). Based on the factor loadings, I designate variables (within each capacity) to an appropriate capacity factor. Per the recommendation of the literature, I employ an *eigenvalue*[6] higher than one as a criterion to retain a factor (Goretzko, Pham, and

---

[5] After conducting pairwise correlations, I find some correlations are higher than others. Overall, most correlations were significant at p=0.05

[6] An eigenvalue is the amount of variance in the sample, which is explained by each factor. The eigenvalue is calculated by summing the squared factor loadings for that factor.

Bühner 2019). I also observe *Scree plots* to assess the number of extracted factors (see Appendix Figures C.1-C.6).[7]

The confirmatory factor analyses lead to the generation of 13 factors (see factors' descriptive summary below in Table 1).[8] I name them such that they capture the essence of underlying variables. Later, I run post-estimation tests, and I find strong evidence for uncorrelation among the factors.[9] In other words, factor analysis returns distinct latent constructs.

**Table 1. Descriptive Statistics of 13 Capacities' Factors Obtained Through CFA**

| Variable | Obs | Mean[10] | Std. Dev.[11] | Min | Max |
|---|---|---|---|---|---|
| **Technology Capacity** | | | | | |
| Base sci & tech | 1230 | 0 | 1 | -.87 | 13.51 |
| Medium sci & tech | 1230 | 0 | 1 | -2.47 | 4.54 |
| High sci & tech | 1230 | 0 | 1 | -1.65 | 6.65 |
| | | | | | |
| **Financial Capacity** | | | | | |
| Financial infrastructure | 1230 | 0 | 1 | -2.68 | 4.86 |
| Financial environment | 1230 | 0 | 1 | -1.29 | 6.69 |
| Strength of financial regulation | 1230 | 0 | 1 | -2.07 | 7.55 |
| Enabling financial environment | 1230 | 0 | 1 | -4.23 | 4.85 |
| | | | | | |
| **Human Capacity** | | | | | |
| Specialized skills | 1230 | 0 | 1 | -2.43 | 2.58 |
| Generalized skills | 1230 | 0 | 1 | -3.48 | 3.01 |
| | | | | | |
| **Infrastructure Capacity** | | | | | |
| Infrastructure (ICT & energy) | 1230 | 0 | 1 | -1.31 | 4.23 |
| Logistic Per. Index (trade & transport infras.) | 1230 | 0 | 1 | -3.36 | 3.45 |
| **Public Policy Capacity** | | | | | |
| Public policy (inc. fiscal, monetary, structural policies…) | 1230 | 0 | 1 | -3.78 | 2.9 |
| | | | | | |
| **Social Capacity** | | | | | |
| Social capacity (inc. equity, inclusion…) | 1230 | 0 | 1 | -3.83 | 2.2 |

[7] Scree plot is a powerful visual tool for determining the number of factors to be retained. It is basically a plot of the eigenvalues shown in decreasing order.

[8] I perform a post-estimation Kaiser-Mayer-Olkin (KMO) test to check the appropriateness of factor analysis to these data (Kaiser 1974). KMO values range between 0 and 1, with small values indicating that the variables have not much in common to warrant factor analysis. Here, by including all variables, the KMO test return a value of about 0.80, suggesting factor analysis is appropriately applied (Watson 2017).

[9] A post-estimation *estat common* displays correlation matrix. Since the factors are orthogonally loaded, the common factors obtained are uncorrelated, as evidenced by identity matrix (STATA Manual).

[10] By definition, the factors are standardized, and their mean values are zero.

[11] Again, by definition the factors are standardized, which means each factor has a SD (and variance) of 1.

Table 1 shows 13 factors extracted from a set of 47 variables. The factors are standardized, and their ranges vary, with technology capacity factors exhibiting the highest range. These 13 capacity factors, their respective capacities, and the specific variables I extract the factors from are shown in Table 2 below.

**Table 2. Capacities (6), Capacity Factors (13), and their Variables (47)**

| Capacity Names | Capacity Factors | Variables |
|---|---|---|
| **Technological Capacity** | 1. Base sci & tech | Sci & tech. articles<br>Intellectual property payments (mil)<br>Voc. & tech. students (mil) |
| | 2. Medium sci & tech | R&D researchers (per mil)<br>ECI (economic complexity) |
| | 3. High sci & tech | R&D expenditure % of GDP<br>R&D technicians (per mil)<br>High-tech exports (mil) |
| **Financial Capacity** | 4. Financial infrastructure | Domestic credit by banks<br>Business density<br>Financial accountholders<br>Commercial banks |
| | 5. Financial (business) environment | Business startup cost<br>Days to start a business<br>Openness measure |
| | 6. Strength of financial regulation | Tax revenue (% of GDP) (tax capacity)<br>Days enforcing a contract<br>Days to register property |
| | 7. Enabling financial environment | Days to obtain electric meter<br>Days to register property (also loads moderately on this variable) |
| **Human Capacity** | 8. Specialized skills | Human Capital Index 0-1<br>Industry employment<br>Service employment<br>Govt. expend. on educ. (loads moderately)<br>Compulsory educ. (years) (loads moderately)<br>Secondary enrollment (gross)<br>Primary completion rate |
| | 9. Generalized skills | Primary enrollment<br>Primary pupil-teacher ratio<br>Advanced education labor (loads very low, though) |
| **Infrastructure Capacity** | 10. Infrastructure (ICT & energy) | Mobile subscriptions<br>Access to electricity<br>Broadband subscriptions<br>Telephone subscriptions |

| Capacity Names | Capacity Factors | Variables |
|---|---|---|
| | | Energy use<br>Internet users |
| | 11. Logistic Per. Index (trade & transp. i~) | Logistic perf. Index 1-5 |
| Public Policy Capacity | 12. Public policy factor (inc. fiscal, monetary, structural policies…) | Statistical capacity 0-100<br>CPIA economic management<br>Public sector management & institutions<br>Structural policies<br>Legal Rights Index 0-12 |
| Social Capacity | 13. Social capacity factor (inc. equity, inclusion…) | Human resources rating<br>Equity of public resource use<br>Social protection rating<br>Social inclusion<br>National headcount poverty (loads low though)<br>Social contributions (loads moderately) |

For the complete detail about the variables, their units, and sources, please refer to Table A.1 in the Appendix. As shown in Table 1 here, in the case of technology capacity, the variables considered are grouped into three factors: *base science and technology* (as reflected in journal articles, payments for intellectual use, and secondary education pupils enrolled in technical and vocational education programs), *medium science and technology* (as indicated by economic complexity score calculated by Harvard's Center for International Development and researchers in R & D and R & D researchers), and *high science and technology* (high technology exports, R & D expenditure, and technicians in R & D). While the first factor indicates a general research culture, the latter two factors generally portray innovation and invention, and they may approximate Kim's concept of "innovation capability."

In the case of financial capacity, factor analysis creates four important factors. The first factor I name is *financial infrastructure* as indicated by account ownership, commercial bank branches, new business density, and domestic credit by the banking sector. The second factor I call is *financial (business) environment* as reflected in days required to start a business, economy openness, and cost of business startup procedures. The third factor is the *strength of the financial regulations* as measured by days to enforce contracts, the ability to collect tax revenue, and the days required to register a property. In contrast, the fourth factor is *an enabling financial* *environment* as indicated by days to obtain an electric connection and days to register a property to a moderate extent.

Regarding human capacity, the analysis generates two broad factors; the first is the *generalized skill level of the population* as reflected in primary enrollment, primary pupil-teacher ratio, alongside the variable labor force with advanced education albeit the factor's low loading on this variable. A second factor refers to the *specialized skill level of a population* as shown by the World Bank's Human Capital Index score, employment in industry and service sectors, secondary enrollment, primary completion rate, government expenditure on education, and compulsory education duration. While the factor loads highly on many variables, it loads moderately on the last two variables.

Similarly, the article extracts two factors for infrastructural/infrastructure capacity. I call *general infrastructure* the first factor, as captured by ICT infrastructure, including broadband subscribers, telephone subscribers, mobile cell subscribers, internet users, and energy infrastructure proxied by per capita energy use and access to electricity. The second factor indicates the *quality of trade and transport-related infrastructure,* including ports, roads, and railways (as proxied by logistic performance index score calculated by the World Bank). In essence, the two factors might be equivalent to Kim's "production" capability coined in the context of firms. Thus, ICT penetration, energy provision, and transport-related infrastructure are crucial for a country's economic progression as they are for firms' ability to produce and market goods and services and compete in international markets.

In light of my confirmatory-led information, the analysis groups variables into two factors based on factor loadings for the last two capacities. Since I conceptualized these two capacities more uniquely, they merit more attention. First, I capture the public policy capacity by a *public policy factor*, which loads highly on variables about public sector management and institutions, economic management, structural policies, the strength of legal rights, and statistical capacity scores of countries. All of these are composite indicators that the World Bank Group constructed based on the data they collected. For instance, *public sector management and institutions* are composite measures, indicating property rights and rule-based governance, quality of budgetary and financial management, the efficiency of revenue mobilization, quality of public administration, transparency, accountability, and corruption in the public sector. Similarly, *structural policies* indicator includes trade and business regulatory environment. On the other hand, the *economic management* indicator measures macroeconomic management, fiscal policy, and debt policy. Furthermore, the *legal rights* index measures the extent to which laws protect the rights of borrowers and lenders. Finally, all these policies require a solid statistical capacity (as measured in the statistical capacity score) in a country to report the findings timely and periodically for policy formulation, coordination, and implementation.

By incorporating all the traditional governance measures (corruption, the rule of law, and accountability in the public sector, business regulatory environment) and other broader measures for governance (fiscal policy, monetary policy, debt policy, macroeconomic management) alongside new measures (statistical capacity and legal rights), the public policy factor is very encompassing. In a way, this factor is a good fusion of neoclassical and traditional capacities approaches: while the most conventional measures included in this factor approximate Abramovitz's social capacity (1986), the broader measures are neoclassical.

Lastly, the analysis captures dimensions of social capacity by a *social capacity factor*. This factor loads highly on policies for social inclusion, human resource rating, social protection rating, equity of public resource use, poverty headcount ratio, and social contributions. Again, some of these are composite indicators constructed by the World Bank. For instance, the social inclusion indicator includes gender equality and policies and institutions for environmental sustainability, among other things. Similarly, social protection rating assesses government policies in social protection and labor market regulations that reduce the risk of becoming poor. Equity of public resource use, on the other hand, evaluates the degree to which public expenditures and revenue collection affects the poor and is consistent with national poverty reduction. Finally, poverty headcount indicates poverty, whereas social contributions are contributions by employees to social insurance schemes operated by the government. At the heart of this capacity is how societal members benefit each other and whether and how the government creates an enabling environment in terms of regulations and social policies to cater to the vulnerable and poor in society. Further information on definitions and sources of all the six capacities variables can be found in Appendix Table A.1.

### 7. Results and Discussion:

Here I report results and discuss some key findings. In doing so, I explain the rationale behind the technical methods and models that I employ in this paper. Some readers who are familiar with them may skip the details and focus on results. However, briefly illustrating why I choose one model or method will benefit readers in general.

To recap, in this article, I estimate the impact of all capacities (as framed in the NACS) on per capita GDP (outcome) using the MSK panel dataset. As illustrated, based on the six capacities, confirmatory factor analyses returned a total of thirteen factors (Table 1). These 13 factors serve as independent variables. My empirical estimation controls confounders that can possibly influence the relationship between economic growth and estimated capacity factors to find the true effect size.

The initial equation (1) mentioned in section 4 bears repeating here.

$$Log\ Y = \Sigma\ \alpha_n C_n\ +\ \Sigma\ \beta_n Z_n\ +\ \varepsilon \tag{1}$$

Log Y is the natural logarithm of GDP per capita, Cn is composed of absorptive capacities factors, Zn is composed of controls, $\alpha_n$ is a vector of parameters that capture the effects of capacities factors on the log of GDP per capita, and $\beta_n$ is a vector of parameters that capture the effects of controls on the log of GDP per capita.

As mentioned, the absorptive capacities factors are independent variables—13 factors (from Table 1). Controls, on the other hand, include *incoming dimensions* (such as technological cooperation grants, international tourist arrivals, merchandize import from high-income countries, net ODA and official assistance received) and other variables (health expenditure as a percentage of GDP, employers' percentage in total employment, total population, and gross capital formation). Since my independent variables are standardized factors, I standardize all the control variables as well; all controls have a mean of zero and SD of 1.

As a preliminary analysis, I estimate a simple pooled OLS regression—an OLS estimation run on panel data (Collischon and Eberl 2020). Pooled OLS returns all statistically significant results, including a positive and significant relationship for public policy factor (see Table 3 below). However, Pooled OLS does not serve our purpose here for two reasons. One Pooled OLS is most suitable when a researcher selects a different sample for each year in the data (Wooldridge 2010). However, here the same sample of countries is observed along different years, warranting a different model. Secondly, applying OLS on panel data is tantamount to ignoring all country-specific effects. This omission leads to a violation of many basic assumptions, including independence of the error term.

To systematically determine the extent to which the data are *poolable* and, subsequently, if Pooled OLS is the correct estimation for these data, I conduct the Breusch-Pagan Lagrange multiplier test (Onali, Ginesti, and Vasilakis 2017).[12] A significant test result confirms unobserved effects (country-specific) across countries, which must be accounted for. The test diagnostic further indicated that Pooled OLS is not the appropriate estimation for these data.

To cater to the problems posed by Pooled OLS modeling, I use *Random* and *Fixed Effects* Models. Such models are employed when the same sample of countries is observed longitudinally (Onali, Ginesti, and Vasilakis 2017).

Random and Fixed Effects models incorporate country-level effects (Bell, Fairbrother, and Jones 2019). However, they use different modeling procedures.[13] With Fixed Effects models, time-invariant variables are held constant or "*fixed*" (Collischon and Eberl 2020). On the other hand, Random Effects estimate the effects of time-invariant variables (Bell, Fairbrother, and Jones 2019). Different considerations affect the choice between the two models. The most important consideration is the issue of omitted variable bias. If the country-level effects (independent of the explanatory variables) are included in the model, then a Random Effects model is preferred. However, if there are omitted variables, and these omitted variables are correlated with explanatory variables, then Fixed Models may control for an omitted variable bias to the extent that it is time-invariant (Kropko and Kubinec 2020).[14]

---

[12] The null hypothesis for this test is that the variance of the unobserved Fixed Effects is zero. A highly significant test results (chisq=4647 and p=0.0001) suggests rejecting the null hypothesis (var=zero for countries). This indicates that Pooled OLS is not an appropriate model for this data.

[13] Fixed Effects models capture *within* variability, whereas Random Effects models capture both *within* and *between* variability. Both models have pros and cons. In general, Random Effects models more often have smaller *standard errors,* but they more likely produce *biased* estimates. On the other hand, Fixed Effects models may produce larger standard errors but more likely unbiased estimates.

[14] Fixed Effects models hold the effects of time-invariant variables (whose values do not change with time, for instance, gender) constant. This means that whatever effects omitted variables have on the subject at one time, they have the same effect on the later time; hence their effect is "constant" or "fixed."

Here, I focus on results from Fixed Effects modeling because Fixed Effects are more meaningful for few reasons. First, Fixed Effects control for time-invariant characteristics of LMICs, which are otherwise hard to incorporate. Secondly, it is impossible to include all the variables that impact economic growth because the countries in this sample have poor data environments. Thirdly, these omitted variables can be correlated with the explanatory variables (capacity factors), leading to biased estimates. Fixed Effects models alleviate these problems. In terms of technical diagnosis, a significant Hausman test also rules in favor of Fixed Effects modeling.[15]

Furthermore, as per the literature recommendation (Wooldridge 2010), I have included dummy variables for years (time effects) in all the models in this article. Similarly, I have incorporated robust standard errors.[16] The inclusion of time dummies and robust errors corrects heteroskedasticity and other inertial effects (Stock and Watson 2008).[17]

Table 3 reports the results of the models discussed above. A complete table including estimates for control variables and year effects is available in Appendix D. Here, I have excluded them for brevity.

---

[15] After conducting Hausman test, I reject null hypothesis that difference in coefficients under the two modeling is not systematic (chi2= 163 and p-value=0.0001). This indicates to conduct Fixed Effects modeling.

[16] I conduct all the regressions reported here in the paper with normal standard errors as well (details in sensitivity analysis). Significances of the factors do not alter; however, standard errors are lower than the robust errors.

[17] Heteroskedasticity is the variance of the error term in a regression model in an explanatory variable, which violates model assumptions. It needs to be diagnosed and corrected. I have used Brusch Pagan (null hypothesis= constant variance) and White tests (null hypothesis= homoskedascity) to detect heteroskedascity. BP returns significant results (chi2= 18 and p=0.0001), suggesting hetroskedasity. Similarly, White test returns significant results (chi2=892 and p=0.0001), again suggesting heteroskedascity.

**Table 3: Main Regressions Results. Dependent Variable, Log of GDP Per Capita.**

| VARIABLES | Pooled OLS | Random Effects | Fixed Effects |
|---|---|---|---|
| Public policy (inc. fiscal, monetary, structural…) | 0.098*** | 0.077*** | 0.087*** |
| | (0.022) | (0.025) | (0.027) |
| (General) Infrastructure (ICT & energy) | 0.371*** | 0.134*** | 0.095*** |
| | (0.026) | (0.024) | (0.028) |
| Logistic Per. Index (trade & transport infrast.) | 0.100*** | 0.037*** | 0.029*** |
| | (0.016) | (0.009) | (0.009) |
| Specialized skills | 0.240*** | 0.111*** | 0.061*** |
| | (0.024) | (0.018) | (0.018) |
| Generalized skills | -0.081*** | -0.030** | -0.018 |
| | (0.012) | (0.015) | (0.014) |
| Financial infrastructure | 0.109*** | 0.037*** | 0.025** |
| | (0.017) | (0.013) | (0.012) |
| Financial (business) environment | 0.047*** | 0.023** | 0.025** |
| | (0.015) | (0.010) | (0.010) |
| Strength of financial regulations | -0.045*** | 0.010 | 0.010 |
| | (0.014) | (0.018) | (0.018) |
| Enabling financial environment | 0.036*** | 0.003 | 0.002 |
| | (0.011) | (0.005) | (0.005) |
| Base sci & tech | -0.179*** | -0.023 | -0.028 |
| | (0.034) | (0.023) | (0.025) |
| Medium sci & tech | -0.127*** | -0.001 | 0.002 |
| | (0.017) | (0.013) | (0.013) |
| High sci & tech | -0.061*** | -0.002 | 0.001 |
| | (0.014) | (0.006) | (0.006) |
| Social capacity (incl. equity, inclusion, etc.) | -0.128*** | 0.004 | -0.001 |
| | (0.023) | (0.019) | (0.018) |
| Constant | 7.329*** | 7.181*** | 7.149*** |
| | (0.051) | (0.068) | (0.032) |
| | | | |
| Observations | 1,230 | 1,230 | 1,230 |
| R-squared | 0.799 | 0.727 | 0.468 |
| Control Variables | YES | YES | YES |
| Year Fixed Effects | YES | YES | YES |
| Country Fixed Effects | NO | NO | YES |
| Robust Standard Errors | YES | YES | YES |
| Number of countries | 82 | 82 | 82 |

Robust standard errors in parentheses
*** p<0.01, ** p<0.05, * p<0.1

The Fixed Effects estimates return six capacities factors to be statistically significant. The public policy capacity factor appears to be significant throughout the models, indicating public policy capacity impacts GDP per capita growth (hereafter, *economic growth*) in LMICs. A one-point standard deviation (SD) increase in the public policy capacity leads to about an 8.7 percent increase in economic growth, holding other factors constant.[18] While the particular magnitude coefficient size for this factor ranged from about 8.7 in the Fixed Effects specification to 9.8 percent in the Pooled OLS model, this is significant across different specifications, including robust errors, Pooled OLS, and Random Effects, and Standard Error models (see Appendix I for further details).

Similarly, the two infrastructure capacity factors exhibit a positive and significant impact on economic growth throughout. The estimated coefficient magnitude of the first infrastructure capacity factor (general) infrastructure (includes ICT and energy infrastructure) ranges from about 9.5 percent in the Fixed Effects model to 37 percent in the Pooled OLS model. For the LPI factor (transport and trade-related infrastructure), the estimated coefficient size ranges from about 2.9 percent to 10 percent. Thus, interpreting the former factor in the Fixed Effects specification means that increasing ICT and energy infrastructure by 1 standard deviation in LMICs increases economic growth by about 9.5 percent, holding other factors constant. Similarly, increasing transport and trade-related infrastructure by 1 standard deviation in LMICs improves economic growth by 2.9 percent, holding other factors constant.

While the generalized skills factor is insignificant in the Fixed Effects specification within the human capital capacity, specialized skills positively and significantly impact economic growth in LMICs. The estimated coefficient size of this particular factor ranges from 6 percent (Fixed Effect) to 24 percent (Pooled OLS). A 6 percent estimated coefficient means that increasing 1 SD of specialized skills factor (which includes HCI, service, and industry sector employment, among other things) in a country increases GDP per capita by 6 percent while holding all other factors constant.

Finally, within financial capacity, I observe two factors of financial infrastructure (including banks, credit, businesses, among others) and environment (including the cost of business start-up procedures, economy openness, and days required to set up a business) as significant. The estimated coefficient size for both these factors is the same, meaning both are equally important. By improving 1 SD of financial infrastructure or 1 SD of the financial (business) environment, GDP per capita in a particular country increases by 2.5 percent.

---

[18] Factors obtained through factor analysis are standardized with a mean of zero and a standard deviation of 1. Therefore, all interpretations in this paper capture the impact of 1 standard deviation change on economic growth. Also, since the dependent variable is the log of GDP per capita, a 1 standard deviation change in any independent variable causes X percent change in the dependent variable.

Overall, trends (directions) in results are very consistent across various specifications in conveying the role of capacities in economic growth. However, magnitudes of effects vary in the three specifications, with Pooled OLS returning higher magnitudes than Random and Fixed Effects. This is a reflection the Pooled OLS models omitted variable bias, which is corrected for in Random and the more conservative Fixed Effects estimates.

Overall, improving public policy capacity and infrastructure capacity factors have positive and significant effects in all cases. However, the results for the financial capacity and human capital capacity factors are more mixed, suggesting that some factors within these capacities are more important than others in LMICs. Specifically, the factor of specialized skills (within human capital capacity), which reflects secondary and vocational school attainment and industry and service sector employment, among other things, is significant and positive throughout. Similarly, financial infrastructure (indicating bank accounts ownership, commercial bank branches, new business density, and domestic credit by banking sector) and financial business environment (measuring days required to start a business, economy openness, and cost of business startup procedures), which appears consistently in the literature, are positive and significant for economic growth in developing economies. All these results corroborate the findings of the existing capacities literature. For instance, in one previous study, the size of the effect of Governance capacity (government effectiveness, corruption, law and order) ranges from 13 percent to 29 percent, whereas the magnitude of Education capacity (primary, secondary, and tertiary attainment) varies from -2 percent to 12 percent (Fagerberg and Serholic 2017). In my study, although the sizes are different, for obvious reasons, because I included an extended set of variables, trends and significances are the same.

On the other hand, in my study, technology, and social capacities, on average, are not significant for economic growth within LMICs. As a matter of fact, the LMICs have not been paying any special attention to these capacities and perhaps rightly so, as these capacities in poor economies by themselves may depend on other capacities, such as foundational public policy, finance, and infrastructure. For example, suppose a poor economy does not have enough fiscal space, strong fiscal management, statistical capacity, and tax capacity as reflected in public policy and financial capacities. In that case, it might not roll out successful social capacity interventions. Similarly, if a developing country's infrastructure (trade infrastructure, electricity provision, and internet communication) is weak, investments in science and technology may hardly lead to any significant economic value. Therefore, while very important for explaining economic growth in High-Income economies, technology and social capacities are seldom priorities of LMICs. Such results are different for LMICs than the existing literature. For example, in the case of technology, compared to insignificant effects in my study, one previous research records large effect sizes ranging from 18 percent to 28 percent for a mix of economies, including higher-income economies (Fagerberg and Srholec 2017). The sizes are likely large because their technology capacity is very narrow: this capacity includes S&T journal articles, R&D expenditures, and USPTO patent applications. Perhaps, their technology capacity is overestimating the impact on economic growth. Regardless of the reason, in my study, economic growth variation is not explained by technology capacity. This finding suggests that in LMICs, technology capacity (including narrow technology indicators of S&T articles, R&D expenditures, Researchers, among others) is not as important.

Different effects sizes from Fixed Effects illustrate that some capacities matter more than others for economic growth in LMICs. For instance, general infrastructure (ICT and energy infrastructure) with an effect size of 9.5 percent tops the list. This factor is followed by the public policy capacity factor with an effect size of 8.7 percent and then the specialized skills capacity factor with an effect size of 6.5 percent. Similarly, the fourth place goes to trade and transport-related infrastructure (effect size 2.9 percent). Finally, the fifth place is captured by financial infrastructure and financial environment, both with an effect size of 2.5 percent. This result about factors' relative importance or ranking is particularly crucial for tight-budget LMICs when prioritizing their investments.

### 7.1 K-Means Cluster Analysis:

To further probe the abovementioned estimates from Fixed Effects models in a disaggregated fashion and glean patterns out of data for policy implications, particularly in significant variables, I decide to further view the data structure. Thus, I first perform cluster analysis using the *Kmeans* algorithm, an unsupervised machine learning tool and analysis approach (Ni et al. 2021). As an iterative algorithm, the *Kmeans* algorithm divides the dataset into k clusters by minimizing intra-cluster variation while keeping the clusters as far as possible. For this analysis, I use a dataset composed of the latest 5-year averages (between 2015 and 2019) because it allows for a study of challenges and policy considerations currently faced by LMICs. To investigate how far countries differ and if they can be clustered in a way that allows us to see a larger picture of the capacity trends in a comparative manner among LMICs, I choose four significant capacity factors (one per each significant capacity) extracted in this paper. These four capacity factors, termed as cluster variables for the k-means analysis, include public policy factor, infrastructure capacity factor showing general infrastructure in ICT and energy, a human capacity factor of specialized skills, and financial capacity infrastructure factor.

To arrive at an optimal k-number (number of clusters) for analysis, I follow a well-established Elbow's approach, explained in Makles (Makles 2012).[19] This approach is primarily based on minimizing With-in cluster Sum of Square (WSS) or intra-cluster variation. Thus, I find evidence for 5 clusters (I have included Scree plots, Appendix Figures F.1 and F.2, and other diagnostic statistics, Appendix Tables F.1 and F.2, for two repeated clusterings).[20]

For the 5-cluster solution obtained as above, I plot a scatterplot matrix (Figure 3 below) of the four standardized factors.

---

[19] When k-numbers are unknown, a researcher repeatedly clusters data using pre-defined number of clusters (Ks). Every time clustering is repeated, intra-cluster variation (i.e., within cluster sum of square or WSS) is recorded. Then the cluster solutions (Ks) are plotted against WSS. From this set of solutions, a researcher chooses the one (optimal k*-cluster solution) that leads to maximum reduction in WSS. At k*, usually the plot shows a kink or elbow. Sometimes, the elbow is not apparent, in which case a researcher keeps on clustering by changing initial conditions (random number). The kink alongside other statistics eventually helps decide in choosing the k* solution. For more details, please check Dr. Anna Makles' work (Makles 2012).

[20] I perform repeated clustering by changing initial condition. In most cases, at k=5, the WSS, log(WSS), and PRE show a kink (plots and statistics in Appendix F, Figures F.1-F.2 and Table F.1-F.2). While the kink in the WSS is less obvious, it is more visible in the log(WSS) and PRE. This naked eye test has to be viewed in conjunction with other statistics. For instance, $\eta2$ shows a reduction of the WSS by 75% and PRE5 to a reduction of about 88%, which is considerably higher reduction when compared with the k = 4 or k = 3 solution. However, the reduction in WSS is very small for k > 5. Where $\eta2$ for a k measures the proportional reduction of the WSS for each cluster solution k compared with the total sum of squares (TSS), PREk illustrates the proportional reduction of the WSS for cluster solution k compared with the previous solution with k − 1 clusters (Makles, 2012)

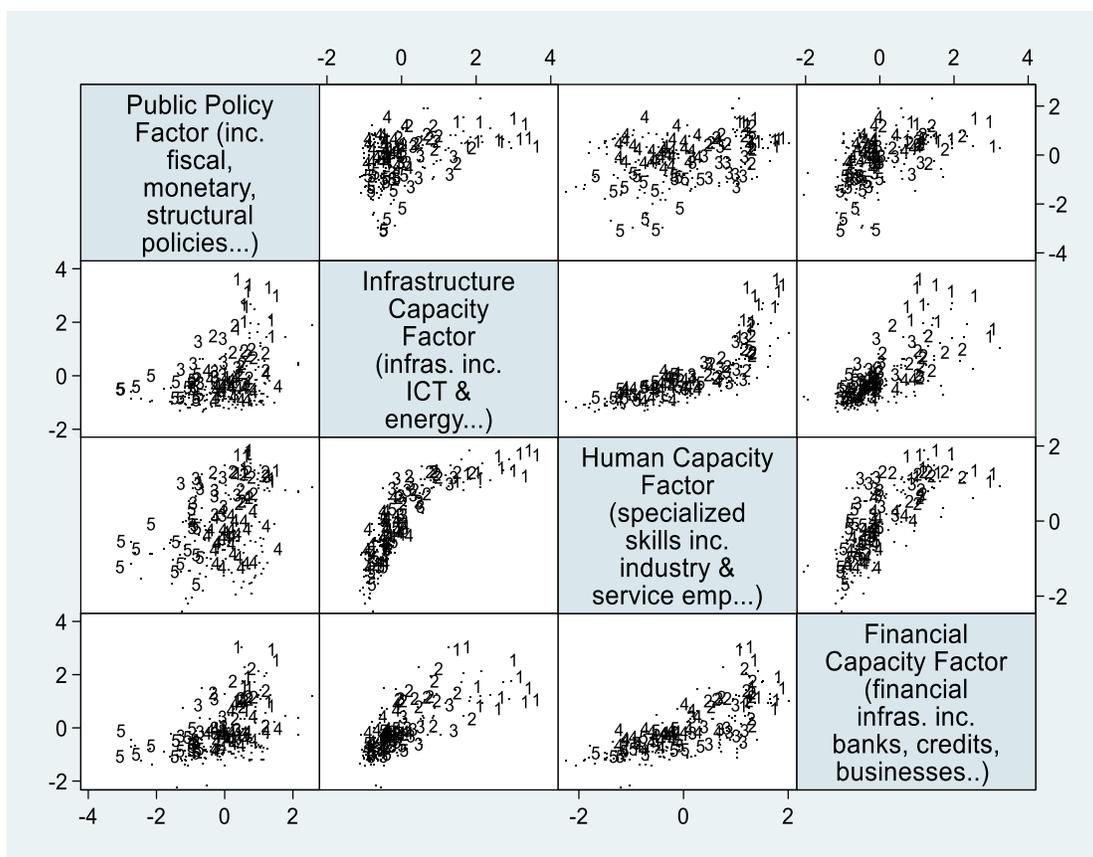

**Figure 3: Scatterplot matrix of four capacity factors for the five-cluster solution. 1s are leaders, 2s are walkers, 3s are creepers, 4s are crawlers, and 5s are sleepers.**

The scatterplot shows five groups of LMICs. Some countries or groups of countries (aka clusters) are doing better than others on the four capacity factors. While naming these clusters, I borrowed some vocabulary from the neurodevelopmental phases of human life. After all, just like humans and their bodies' neurodevelopment and thought progression, these countries are also undergoing development phases in their lifetime as part of the global world. Hence, the names of the clusters are *leading, walking, creeping, crawling,* and *sleeping*.[21] High-developed LMICs are in the leading and walking clusters, whereas lower developed LMICs are in the sleeping and crawling groups. Creeping economies are the limping middle-category countries trying to catch up with the walking and leading economies. These clusters and their respective countries are shown below in Table 4 below.

---

[21] The naming order shows a progression such that **leading>walking>creeping>crawling>sleeping**. All categories are intuitive. However, readers may rightly be confused about creeping and crawling as they are used interchangeably in everyday life. However, they are distinct neurodevelopmental stages. Creeping is level 3 in mobility after crawling, which is level 2 (please see this link: https://www.domaninternational.org/blog/creeping-a-vital-developmental-stage). In other words, creeping is higher than crawling. In crawling, the body is in contact with the surface and in creeping it is raised above the floor (McGraw 1941).

**Table 4: LMICs Divided into 4 Clusters Following a *Cluster Analysis* (*K-means*)**

| Clusters | No. of countries | Countries (2015-2019) |
|---|---|---|
| 1. **Leading** | 11 | Mongolia<br>Kosovo<br>Moldova<br>St. Vincent and the Grenadines<br>Grenada<br>Bosnia and Herzegovina<br>Georgia<br>St. Lucia<br>Dominica<br>Vietnam<br>Armenia |
| 2. **Walking** | 13 | Tonga<br>Sri Lanka<br>Uzbekistan<br>Bhutan<br>Nepal<br>Samoa<br>Cabo Verde<br>Bolivia<br>Cambodia<br>Kyrgyz Republic<br>India<br>Maldives<br>Honduras |
| 3. **Creeping** | 12 | Timor-Leste<br>Myanmar<br>Marshall Islands<br>Micronesia, Fed. Sts.<br>Tuvalu<br>Tajikistan<br>Guyana<br>Sao Tome and Principe<br>Bangladesh<br>Nicaragua<br>Kiribati<br>Lao PDR |
| 4. **Crawling** | 30 | Togo<br>Sierra Leone<br>Solomon Islands<br>Djibouti<br>Nigeria<br>Senegal |

| Clusters | No. of countries | Countries (2015-2019) |
|---|---|---|
| | | Zambia |
| | | Kenya |
| | | Malawi |
| | | Lesotho |
| | | Tanzania |
| | | Mali |
| | | Gambia, The |
| | | Mozambique |
| | | Papua New Guinea |
| | | Liberia |
| | | Madagascar |
| | | Mauritania |
| | | Burkina Faso |
| | | Vanuatu |
| | | Ghana |
| | | Guinea |
| | | Cote d'Ivoire |
| | | Pakistan |
| | | Uganda |
| | | Benin |
| | | Ethiopia |
| | | Cameroon |
| | | Rwanda |
| | | Niger |
| **5. Sleeping** | 16 | Congo, Dem. Rep. |
| | | Chad |
| | | Angola |
| | | Zimbabwe |
| | | Haiti |
| | | Burundi |
| | | Congo, Rep. |
| | | Yemen, Rep. |
| | | Afghanistan |
| | | South Sudan |
| | | Guinea-Bissau |
| | | Eritrea |
| | | Comoros |
| | | Somalia |
| | | Central African Republic |
| | | Sudan |

The table shows that many countries (30) are in the crawling cluster while the least number of countries (11) is in the leading cluster. In the next step, I calculate the mean scores of the select four factors for all the clusters listed in Table 4. Results of Cluster K-means by clusters are shown below in Table 5 (detailed descriptive statistics for each cluster are in the Appendix E).

**Table 5. Mean Scores of Select Four Factors and Economic Growth in Five Clusters.**

| Clusters | Public Policy Factor (inc. fiscal, monetary, structural policies...) | Infrastructure Capacity Factor (infras. inc. ICT & energy) | Human Capacity Factor (specialized skills inc. industry & service employment) | Financial Capacity Factor (financial infras. inc. banks, credits, business...) | Total Score by each cluster | Log of GDP Per Capita | Log of GDP Per Capita Growth |
|---|---|---|---|---|---|---|---|
| **Leading (11)** | 0.855 | 2.6503 | 1.451 | 1.667 | 6.624 | 8.5 | 3.16 |
| **Walking (13)** | 0.590 | 0.731 | 1.030 | 1.118 | 3.469 | 7.81 | 3.13 |
| **Creeping (12)** | -0.431 | 0.334 | 0.719 | 0.047 | 0.670 | 7.51 | 1.01 |
| **Crawling (30)** | 0.248 | -0.397 | -0.418 | -0.220 | -0.786 | 6.94 | 0.844 |
| **Sleeping (16)** | -1.444 | -0.447 | -0.619 | -0.660 | -3.170 | 6.8 | -1.29 |
| **Total (82)** | -0.045 | 0.288 | 0.189 | 0.198 | 1.361 | 7.24 | 2.13 |

The first cluster is composed of 16 *sleeping* economies from East Africa, North and Central Africa, and other regions, which primarily suffer from severe internal conflicts and disasters, with very poorly developed public policy capacity, general infrastructure, financial apparatus, and human capital and social capacities. The second cluster, consisting of 30 countries from Africa and a South Asian country (Pakistan), distinguishes itself from the sleeping cluster by having relatively better public policy capacity, some specialized skills, and financial apparatus. While their public policy capacity is better than that of sleeping economies, their general infrastructure (energy and ICT infrastructure) is roughly the same. Since they have somewhat better capacities than sleeping economies (score of -0.786>score of -3.270), they are "awake" and in a developmental phase.

The "*creeping*" cluster consists of 12 economies, including South American (Guyana), Central American (Nicaragua), former Soviet republics (Tajikistan), South Asian (Bangladesh), Southeast Asian (Lao PDR, Myanmar, Timor-Leste), and small oceanic countries (the Marshall Islands and Tuvalu, among others). Their ICT and energy infrastructure, human capacity factor, and finance infrastructure are higher than that of the crawling and sleeping economies. However, their public policy factor is not fully developed, being higher than that of the sleeping countries but lower than that of the crawling countries. This result is a little surprising. Despite the fact that their public policy score is overall low (and lower than crawlers), creepers are still in a higher developmental stage (above the crawlers), as evidenced in their higher economic growth. I explain this partly by advancements in other capacities: because the creepers have higher finance, infrastructure, and human capacities than the crawlers, their overall development and economic growth are higher than the crawlers. To further understand this somewhat surprising result, it helps to draw an analogy from neurodevelopment. Just like the deliberate progressive movement of the older baby whose overall neural organization is of higher-order is distinct from the newborn infants' crawling movements, creeping economies with an average higher capacities score are "developing" distinctly than the crawling economies. This result also demonstrates the fact that countries do not develop linearly. Some countries have better infrastructure and finance; others have strong public policy and human capacity factors. While capacities impact economic growth and their average effect varies in different countries, overall, a country is developing more if it enhances all or most of its capacities.

Then there is a group of 12 economies, which includes former Soviet Republics, South American countries, and South Asian countries, including India and Sri Lanka. They are better than all the above cases. They have a reasonably good public policy, finance, infrastructure, and specialized skills. Since they are certainly in better shape than creeping and crawling, I call them "walking" economies for the sake of this analysis.

 Lastly, there is a "*leading*" cluster of 11 economies from East Europe, East Asia, and the Caribbean. These economies are the best of the lot, with much more advanced capacitiess than elsewhere.

I have calculated the mean of the log of GDP per capita (and its growth) for each group (last two columns of Table 5). The data show leading economies have the highest economic growth, followed by walking, and then creeping, crawling, and lastly, sleeping economies.

This simple analysis bears evidence that countries with higher capacities have higher economic growths. I further demonstrate this in the following scatterplots of significant capacity factors plotted against the log of per capita GDP. All the scatterplots include linear fit lines and shaded grey areas to indicate 95% CI. Figure 4 - Figure 7 plots individual factors against the economic growth, whereas Figure 8 plots a capacity factor index against GDP per capita. Again, I constitute the index by simply taking the average of the four significant factors.

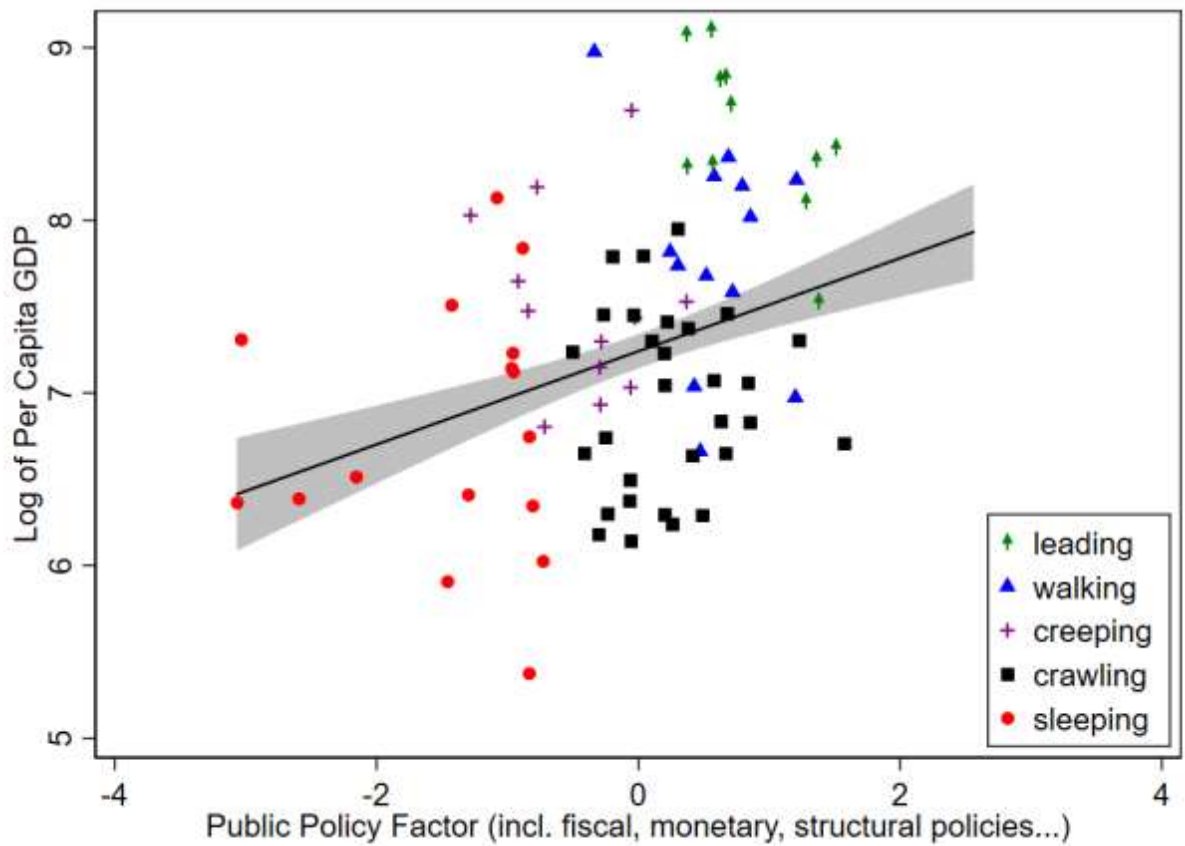

**Figure 4: Public Policy Capacity and Economic Growth in LMICs**

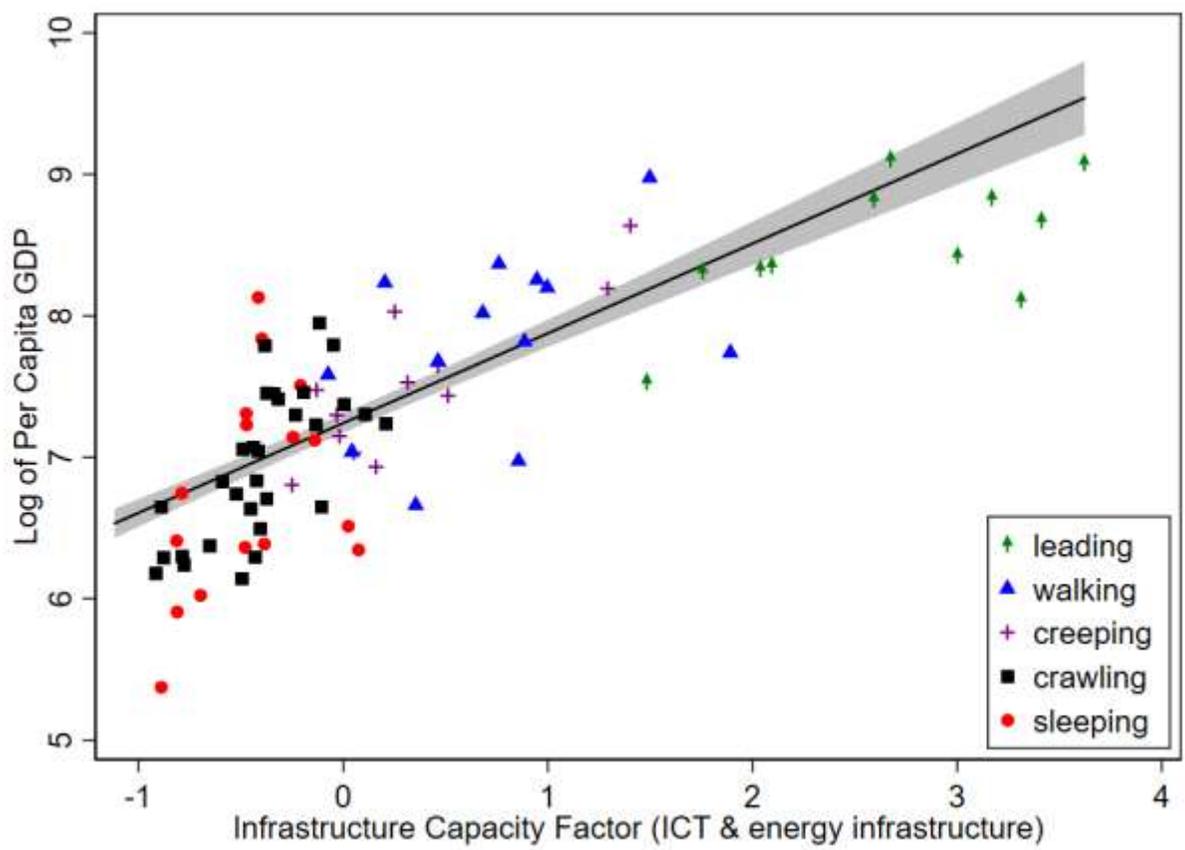

**Figure 5: Infrastructure Capacity and Economic Growth in LMICs**

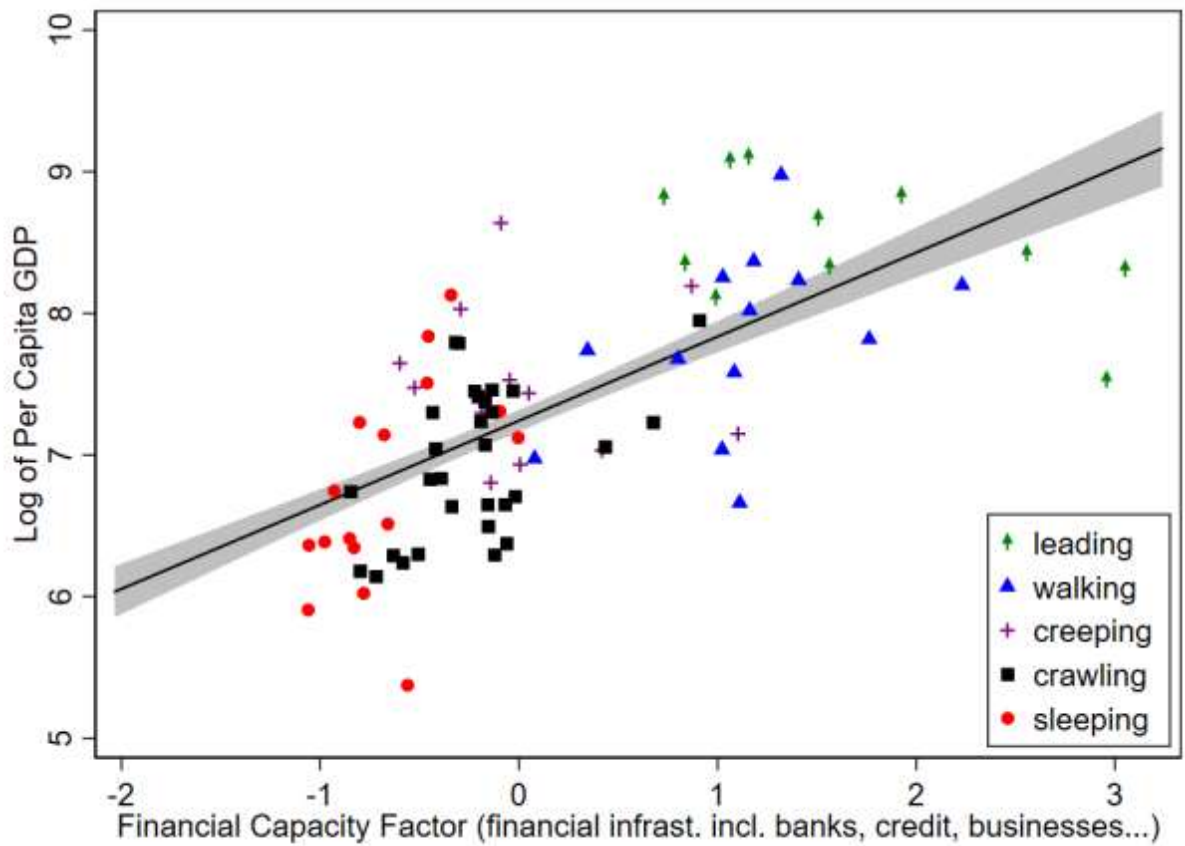

**Figure 6: Financial Capacity and Economic Growth in LMICs**

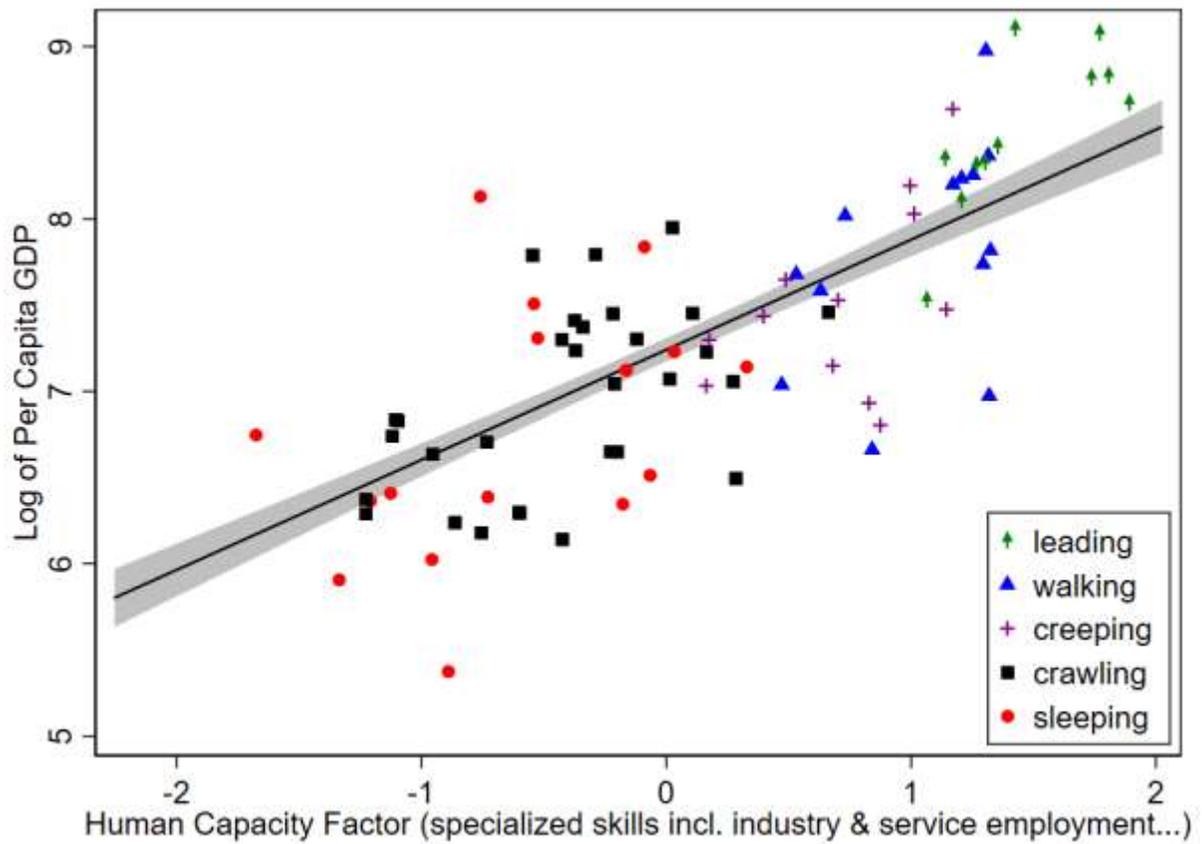

**Figure 7: Human Capital Capacity and Economic Growth in LMICs**

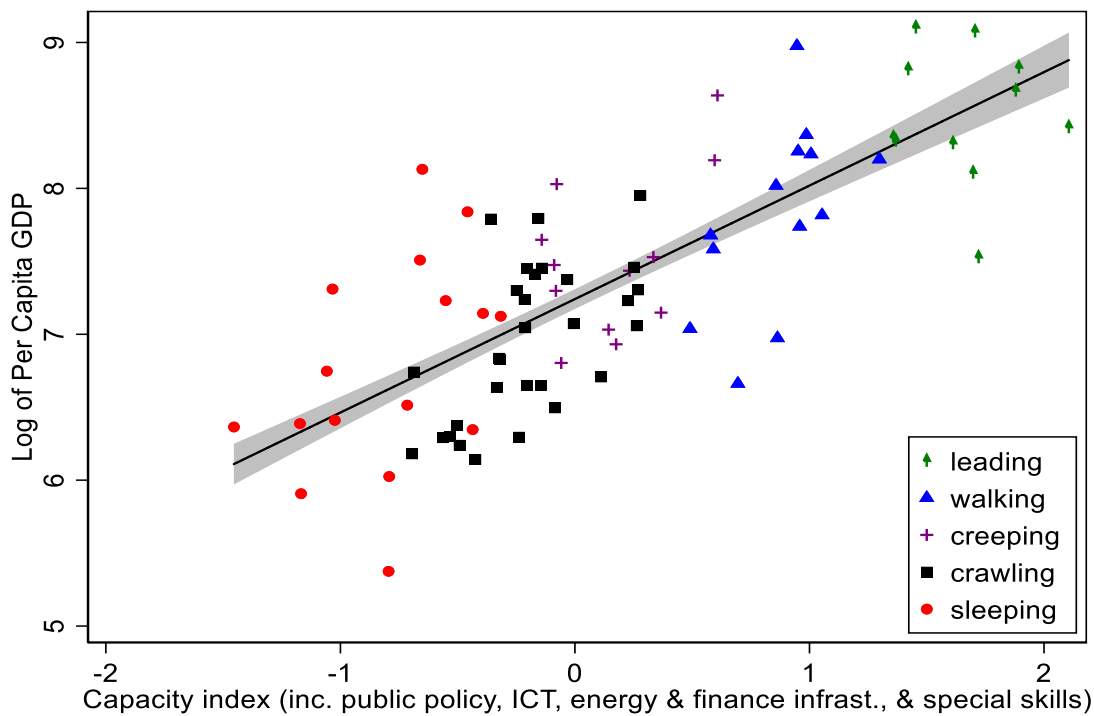

**Figure 8: Capacity factors Index and Economic Growth in LMICs**

The above scatter plots further visually demonstrate the effects of positive and significant capacities on economic growth in LMICs. On one end, there are leading and walking economies. Leading economies are developing their capacities. Walking economies are following the lead of leading economies. On the other end, there are sleeping and crawling economies. While crawling economies are showing some signs of development, sleeping economies are not yet able to overcome the many hurdles and challenges they face. Finally, right in between these two ends are creeping economies. Creeping economies are higher in the hierarchy than crawling. It seems like they are on track to achieve better capacities; however, they still need to put in an extensive effort to catch up with walking economies.

Putting data from plots and Fixed Effects results side-by-side, it is evident that on a granular level, some capacities matter more for economic growth in LMICs (see Section 7 on panel analysis). Generally speaking, countries that have overall higher capacity scores certainly show higher economic growth. With higher capacity and economic growth, leading countries provide examples for the other economies on how they advanced in the development ladder.

A striking observation from my analysis is that countries follow different development patterns even if they are at the same level of development. While, on average, the data and figures point to a gradient in terms of economic growth and significant capacity factors, some countries are walking but still at par with leading countries at least for some capacity factors (see diffused markers or alternating markers on the scatter plots, Figure 5- Figure 9). Similarly, some crawling countries are at par with creeping economies, and likewise, creeping economies are at par with walking economies (again, notice diffused markers on the scatter plots, Figure 5-Figure 9). This is not to say that there are not five distinct groups. Instead, some groups are pooled together for some capacity factors while overall they differ in some other capacity factors, thus leading to five separate and distinct groups.

Nonetheless, the diffused markers may demonstrate two things: firstly, many countries like babies in the development phase crawl first and then creep; however, a few countries creep first and then crawl later, and some creep and crawl together. In other words, creeping and crawling, which manifests itself in economic growth, is impacted by various capacities in different ways. This may also mean that some countries have relatively advanced infrastructure than public policy and vice versa. Others have more advanced finance infrastructure than specialized skills and vice versa. Because of this relative strength of various many capacity factors, diffusion in groups is observed. This observation, in turn, calls for heterogeneity in policy when dealing with or building capacities of LMICs.

Since the average economic growth differences between the clusters (particularly those nearer to each other) are not large, the second observation, which is more of a hypothesis, is that some economies may have been transitioning from one group to another. A cross-country movement (transition analysis) from one group to another over time that incorporates the capacities as I have demonstrated here certainly makes a case for an interesting study. Such a study will reveal whether capacities help countries transition (or graduate) from one development stage to another. This analysis will also indicate whether or not countries are getting better in terms of developing these capacities.

## 8. Sensitivity Analysis:

Here I conduct sensitivity analysis of average treatment effects to examine their robustness. As a first sensitivity analysis, I conduct an exploratory factor analysis (EFA) and then include EFA factors in the same regression models as I show in Table 3, which include factors from confirmatory factor analyses (CFAs). Regressions incorporating EFA factors substantiate the results from regressions, including CFAs factors as predictors. The EFA provides 12 factors (see factor loadings in Appendix G). I retain the factors using the same Eigenvalue criteria I have for CFAs. Then, I include the retained factors in a different set of regressions using the exact specifications of Pooled OLS, Random Effects, and Fixed Effects. Results from the EFA and subsequent regressions are included in Appendix H. By looking at the results from the Fixed Effects specification, the same six factors appear to be significant. Also, the sizes and directions are comparable. For instance, the public policy factor (although now merged with the social capacity factor) obtained from the EFA is significant in all the regressions. The estimated coefficient size of the public policy factor ranges from a comparable 7.6 percent (Pooled OLS) to 11.5 percent (Random). For a Fixed Effects specification, its size is 9.8 percent, which is almost similar to what I observe in the accordingly similar Fixed Effects specification conducted above after including factors from the CFAs.

For an additional sensitivity analysis, I conduct a similar analysis with different specifications. For instance, I run all the above regressions (in Table 3) with standard errors (see Appendix I) while admitting that robust errors are superior and suitable for these models. My main results (estimates size, direction, and their significance) do not alter. Similarly, I perform backward and forward regression analyses with and without time effects (see Appendix J). Despite a drop in estimate sizes, my main results (directions and their significance) do not alter. The drop in coefficient sizes suggests that the use of time effects (in Table 3) is accurate. Similarly, I also perform the same analyses with and without controls (see Appendix J). While the directions and significances do not alter in most cases, some earlier significant capacity factors become insignificant and even change signs.[22] Moreover, estimates sizes in some instances drop considerably with the addition of controls,[23] suggesting that controls must be included in the regression analyses. The change in results (decline in magnitude or significance) with the addition of controls underpins one of my research premises that control variables may influence the relationship. Lastly, I average the entire data over five years (2005-10, 2010-15, 2015-19) and then conduct the same analyses (see Appendix K). My main results, by and large, remain the same.

---

[22] *Base sci & tech* has coefficient estimate of 3.7 percent without controls. After controls are included, it becomes insignificant and size drops to -2.9 percent.

[23] For instance, *social capacity factor* estimate drops from 0.019 to 0.007. Similarly, *strength of financial regulations* coefficient estimate drops from 0.032 to 0.019.

## 9. Conclusion and Implications:

By extending firm-level capacities to a national level, I argue that capacities play a role in the economic development of LMICs. Like all other countries, LMICs are reservoirs of capacities in the form of skills, policies, institutions, resources, and humans. Such capacities engage in interactions to generate economic value. Therefore, capacities need to be integrated into LMICs' economic growth and policy frameworks.

In this article, I develop a framework, National Absorptive Capacity System (NACS), by including four more capacities in addition to the two social and technological capacities in the extant literature and consider important confounders such as incoming flows. I build these capacities utilizing more updated data containing an extensive set of variables. In line with previous research, I find that infrastructure, public policy, finance, business environment, and specialized skills, including service and industry-level employment, are fundamental in explaining economic growth within LMICs. On the other hand, in contrast to existing research for developed countries, I observe technological and social capacity are not significant within LMICs. Perhaps, these capacities will become important later in creating economic value, as LMICs develop a base level of policy, infrastructure, and finance.

For technology capacity, the insignificant result may also suggest that instead of spending recklessly on developing technological base, LMICs, particularly the poorest economies in the first stage, can be more strategic by learning and incorporating from other richer countries. As they rise further on the development ladder, they may start improvising and building their own technological capacities. Thus, I propose richer countries facilitate technology transfer to poorer economies to boost shared prosperity.

Infrastructure capacity, particularly ICT infrastructure and energy provision and infrastructure, stand out as the most crucial capacities for economic growth in LMICs. Similarly, public policy capacity, which indicates fiscal and financial management, quality of institutions, and other bureaucratic reforms, are also extremely important in LMICs. Moreover, these capacities are crucial, as they may have spillover effects on other capacities. That is to say, if an LMIC has a strong infrastructure and public policy capacities, it probably will ensure an enabling environment for various other capacities, thus leading to higher economic value.

My findings have implications for LMICs. By establishing a rich and novel capacities framework to explain economic growth in LMICs and testing the framework on the panel data, I show which capacities and the extent to which those capacities drive economic growth. Policymakers in LMICs can learn from these findings and apply them when devising national growth frameworks. Similarly, other stakeholders and international organizations, such as the World Bank Group, the IMF, and the United Nations, can devise informed strategies in building country-wide growth diagnostic tools and growth partnership frameworks in light of these findings. The cluster analyses, in turn, offer interesting insights and caution against handling LMICs with a one size fits all approach. Overall, in the long run, the findings will help achieve sustainable development goals and shared prosperity in the LMICs, which are prime candidates for development. Future work may look for transition with respect to capacities and economic development within LMICs over time.

**Appendix A Table A.1. Variables Included in NACS Framework, Definitions, and Sources.**

| Capacity | Variable code | Definition | Source |
|---|---|---|---|
| **Technology Capacity** | tippay | **Charges for the use of intellectual property, payments (BoP, current US$).** Payment or charges per authorized use of patents, trademarks, copyrights, industrial processes and designs including trade secrets, and franchises and for the use, through licensing agreements, of produced originals of prototypes. Data are in current US dollars. | IMF, World Bank |
| | tscitjar | **Scientific and technical journal articles.** Number of scientific and engineering articles published in physics, biology, chemistry, mathematics, clinical medicine, biomedical research, engineering and technology, and earth and space sciences, per million people. | World Bank |
| | trandd | **Research and development expenditure (% of GDP)** | UNESCO |
| | tresinrandd | **Researchers in R&D (per million people)** | UNESCO |
| | ttechinrandd | **Technicians in R&D (per million people)** | UNESCO |
| | thigexperofmanex | **High-technology exports (% of manufactured exports).** High-technology exports are products with high R&D intensity, such as in aerospace, computers, pharmaceuticals, scientific instruments, and electrical machinery. | UN, COMTRADE |
| | tsecedvoc | **Secondary education, vocational pupils.** Secondary students enrolled in technical and vocational education programs, including teacher training. | UNESCO |
| | teciscore | **ECI Score.** Measure of economic complexity containing information about both the diversity of a country's export and their sophistication. High ECI Score shows that an economy exports many goods that are of low ubiquity and that are produced by highly diversified countries. Diverse and sophisticated economies have high scores. | OEC, MIT |

| Capacity | Variable code | Definition | Source |
|---|---|---|---|
| **Financial Capacity** | fdaystoenfctt | **Time required to enforce a contract (days).** Days required to enforce a contract, whereas the days are counted from the day a plaintiff files the lawsuit in court until payment. Low values indicate high competitiveness and vice verca. | World Bank, Doing Business Project |
| | fdomcrprsecbybkpergdp | **Domestic Credit by Banking Sector.** This includes all credit to various sectors (monetary authorities, banks, financial corporations) on a gross basis, with the exception of credit to the central government, which is net, as a percentage of GDP. | IMF, World Bank |
| | fopenind | **Openness Indicator.** (Import + Export)/GDP. Constant US 2010. | World Bank |
| | fdaystoregpro | **Time required to register property (days).** The number of calendar days needed for businesses to secure rights to property. | World Bank, Doing Business Project |
| | fcosbstpropergni | **Cost of business start-up procedures (% of GNI per capita)** | |
| | ftaxrpergdp | **Tax revenue (% of GDP).** Tax revenue means compulsory transfers to the government for public purposes. | IMF, WBG |
| | fcombkbr1k | **Commercial bank branches (per 100,000 adults)** | IMF, World Bank |
| | fdaystoobtelecconn | **Time to obtain electrical connection (Days).** Days to obtain electrical connection. Days from application to getting the connection. | World Bank, Enterprise Survey |
| | ftdaystobusi | **Time required to start a business (Days).** The number of days needed to complete the procedures to legally operate a business. | World Bank, Doing Business Project |
| | faccownperofpop15p | **Account ownership at a financial institution or with a mobile-money-service provider (% of pop ages 15+).** Account denotes the percentage of respondents who report having an account at a bank or another type of financial institution or report personally using a mobile money service in the past 12 months (% age 15+). | Demirguc-Kunt et al., 2018, Global Financial Inclusion Database, World Bank. |

| | fnewbusdenper1k | **New business density (new registrations per 1,000 people ages 15-64)**. New businesses registered are the number of new limited liability corporations registered in the calendar year. | World Bank, Enterprise Survey |
|---|---|---|---|
| | | | |

| Capacity | Variable code | Definition | Source |
|---|---|---|---|
| **Human Capacity** | hprimenrollpergross | **School enrollment, primary (% gross).** Ratio of total enrollment, regardless of age, to the population of the age group that officially corresponds to the primary level. | UNESCO |
| | hsecenrollpergross | **School enrollment, secondary (% gross).** Ratio of total enrollment, regardless of age, to the population of the age group that officially corresponds to the secondary level. | UNESCO |
| | hcompeduyears | **Compulsory education, duration (years).** Duration of compulsory education is the number of years that children are legally obliged to attend school. | UNESCO |
| | hgvtexpedupergdp | **Government expenditure on education (% of GDP).** General government expenditure on education (current, capital, and transfers) is expressed as a percentage of GDP. | UNESCO |
| | hpupteapriratio | **Primary pupil-teacher ratio**. Ratio (number of pupils enrolled in primary school) / (number of primary school teachers) | UNESCO |
| | hempinduspertotem | **Employment in industry (% of total employment).** Industry sector comprise mining and quarrying, manufacturing, construction, & public utilities (electricity, gas, & water), | ILO, World Bank |
| | hempserpertotem | **Employment in services (% of total employment).** The services sector consists of wholesale and retail trade and restaurants and hotels; transport, storage, and communications; financing, insurance, real estate, and business services; and community, social, and personal services. | ILO, World Bank |
| | hprimcompra | **Primary completion rate, total (% of relevant age group)** | UNESCO |
| | hhciscale0to1 | **Human capital index (HCI) (scale 0-1).** The HCI calculates the contributions of health and education to worker productivity. The final index score ranges from zero to one and measures the productivity as a future worker of child born today relative to the benchmark of full health and complete education. | World Bank |
| | hlfwithadedu | **Labor force with advanced education (% of total working-age population with advanced education)** | ILO, World Bank |
| | | | |

| Capacity | Variable code | Definition | Source |
|---|---|---|---|
| **Infrastructure Capacity** | imobsubper100 | **Mobile cellular subscriptions (per 100 people).** | International Telecom Union, World Bank |
| | itelesubper100 | **Fixed telephone subscriptions (per 100 people)** | International Telecom Union, World Bank |
| | ibdbandsubper100 | **Fixed broadband subscriptions (per 100 people)** | International Telecom Union, World Bank |
| | iaccesselecperpop | **Access to electricity (% of population).** The percentage of population with access to electricity. | World Bank, Sustainable Energy for All |
| | ienergyusepercap | **Energy use (kg of oil equivalent per capita).** The use of primary energy before transformation to other end-use fuels, which is equal to indigenous production plus imports and stock changes, minus exports and fuels supplied to ships and aircraft engaged in international transport. | IEA, World Bank |
| | iindintperpop | **Individuals using the internet (% of population).** Internet users are individuals who have used the Internet (from any location) in the last 3 months. | International Telecom Union, World Bank |
| | ilpiquoftratraninfr | **Logistics performance index: Quality of trade and transport-related infrastructure (1=low to 5=high).** Logistics professionals' perception of country's quality of trade and transport related infrastructure (e.g. ports, railroads, roads, information technology), on a rating ranging from 1 (very low) to 5 (very high). Scores are averaged across all respondents. | World Bank |
| | | | |
| | | | |

| Capacity | Variable code | Definition | Source |
|---|---|---|---|
| | pcpiapsmgandinscl1to6 | **CPIA public sector management and institutions cluster average (1=low to 6=high).** The public sector management and institutions cluster includes property rights and rule-based governance, quality of budgetary and financial management, efficiency of revenue mobilization, quality of public administration, and transparency, accountability, and corruption in the public sector. | World Bank, CPIA Database |
| **Public Policy Capacity** | pcpiastpolclavg1to6 | **CPIA structural policies cluster average (1=low to 6=high).** The structural policies cluster includes trade, financial sector, and business regulatory environment | World Bank, CPIA Database |
| | pstrengthoflegalright | **Strength of legal rights index (0=weak to 12=strong).** Strength of legal rights index measures the degree to which collateral and bankruptcy laws protect the rights of borrowers and lenders and thus facilitate lending. The index ranges from 0 to 12, with higher scores indicating that these laws are better designed to expand access to credit. | World Bank, Doing Buisness Project |
| | iscapscoravg | **Overall level of statistical capacity (scale 0 - 100).** A composite score (on a scale of 0-100) which assesses the capacity of a country's statistical system in three areas (25 criteria): methodology; data sources; and periodicity and timeliness. | World Bank |
| | pcpiaeconmgtcl1to6 | **CPIA economic management cluster average (1=low to 6=high).** The economic management cluster includes macroeconomic management, fiscal policy, & debt policy. | World Bank, CPIA Database |

| Capacity | Variable code | Definition | Source |
|---|---|---|---|
| | scpiabdhumanres1to6 | **CPIA building human resources rating (1=low to 6=high).** Building human resources assesses the national policies and public and private sector service delivery that affect the access to and quality of health and education services, including prevention and treatment of HIV/AIDS, tuberculosis, and malaria. | World Bank, CPIA Database |
| **Social Capacity** | scpiaeqofpbresuse1to6 | **CPIA equity of public resource use rating (1=low to 6=high).** Equity of public resource use assesses the extent to which the pattern of public expenditures and revenue collection affects the poor and is consistent with national poverty reduction priorities | World Bank, CPIA Database |
| | scpiasocprorat1to6 | **CPIA social protection rating (1=low to 6=high).** Social protection and labor assess government policies in social protection and labor market regulations that reduce the risk of becoming poor, assist those who are poor to better manage further risks, and ensure a minimal level of welfare to all people. | World Bank, CPIA Database |
| | scpiapolsocinclcl1to6 | **CPIA policies for social inclusion/equity cluster average (1=low to 6=high).** The policies for social inclusion and equity cluster includes gender equality, equity of public resource use, building human resources, social protection and labor, and policies and institutions for environmental sustainability | World Bank, CPIA Database |
| | spovheadcnational | **Poverty headcount ratio at national poverty lines (% of population).** National poverty headcount ratio is the percentage of the population living below the national poverty line(s) | World Bank |
| | ssocialconperofrev | **Social contributions (% of revenue).** Social contributions include social security contributions by employees, employers, and self-employed individuals, and other contributions whose source cannot be determined. They also include actual or imputed contributions to social insurance schemes operated by governments | IMF, World Bank |

| Controls | Variable code | Definition | Source |
|---|---|---|---|
| | cteccopgrantbopcurr | **Technical cooperation grants (BoP, Current US $).** Technical cooperation grants are free-standing grants to finance the transfer of technical and managerial skills or of technology with the aim to build national capacity without reference to any specific investment projects; and investment-related technical cooperation grants, which are provided to strengthen the capacity to execute specific investment projects. Data are in current U.S. dollars. | World Bank, International Debt Statistics, and OECD. |
| | cpoptot | **Population, Total.** Total population, counting all residents regardless of legal status or citizenship. | United Nations Statistical Division. |
| | cgroscapformcons2010us | **Gross capital formation (% of GDP).** Gross capital formation consists of outlays on additions to the fixed assets of the economy plus net changes in the level of inventories. Fixed assets include land improvements (fences, ditches, drains, and so on); plant, machinery, and equipment purchases; and the construction of roads, railways, and the like, including schools, offices, hospitals, private residential dwellings, and commercial | World Bank national accounts data, and OECD |
| | | and industrial buildings. Inventories are stocks of goods held by firms to meet temporary or unexpected fluctuations in production or sales, and "work in progress." According to the 1993 SNA, net acquisitions of valuables are also considered capital formation. | National Accounts data files. |
| | cintltournoarrivals | **International tourism, number of arrivals**. International inbound tourists (overnight visitors) are the number of tourists who travel to a country other than that in which they usually reside, and outside their usual environment, for a period not exceeding 12 months and whose main purpose in visiting is other than an activity remunerated from within the country visited. | World Bank |
| | cmerchimpfmhighperm erchimp | **Merchandise imports from high-income economies (% of total merchandise imports).** Merchandise imports from high-income economies are the sum of merchandise imports by the reporting economy from high-income economies according to the World Bank classification of economies. Data are expressed as a percentage of total merchandise imports by the economy. | World Bank |
| | cnetodaandoffreceivcon 2018 | **Net official development assistance and official aid received (constant 2018 US$).** Net official development assistance (ODA) consists of disbursements of loans made on concessional terms (net of repayments of principal) and grants by official agencies of the members of the Development Assistance Committee (DAC), by multilateral institutions, and by non-DAC countries to promote economic development and welfare in countries and territories in the DAC list of ODA recipients. It includes loans with a grant element of at least 25 percent (calculated at a rate of discount of 10 percent). Net official aid refers to aid flows (net of repayments) from official donors to countries and territories in part II of the DAC list of recipients: more advanced countries of Central and Eastern Europe, the countries of the former Soviet Union, and certain advanced developing countries and territories. Official aid is provided under terms and conditions similar to those for ODA. Part II of the DAC List was abolished in 2005. The collection of data on official aid and other resource flows to Part II countries ended with 2004 data. Data are in constant 2018 U.S. dollars. | OECD, World Bank |
| | hcurhealthexppergdp | **Current health expenditure (% of GDP).** Level of current health expenditure expressed as a percentage of GDP.  Estimates of current health expenditures include healthcare goods and services consumed during each year. Does not include capital health expenditures such as buildings, machinery, IT and stocks of vaccines for emergency or outbreaks. | WHO, World Bank |
| | hemplyrpertotemp | **Employers, total (% of total employment).** Employers are those workers who, working on their own account or with one or a few partners. Self-employment included. | ILO, World Bank |
| | timecode | **Time Frame.** from 2005 to 2019 | |
| | countrycode | **Countrycodes** as used by the World Banks | |
| | | | |
| **Outcome** | **Variable code** | **Definition** | **Source** |
| | ogdppercapconst2010us | **GDP per capita (constant 2010 US$).** GDP per capita is gross domestic product divided by midyear population.  Data are in constant 2010 U.S. dollars. | World Bank |

**Appendix A. Table A.2. Descriptive Statistics of All Variables in NACS**

| Variable | Obs | Mean | Std. Dev. | Min | Max |
|---|---|---|---|---|---|
| Sci & tech. articles | 1230 | 1270.77 | 9395.79 | 0 | 135787.8 |
| Intellectual payments (mil) | 1230 | 65.35 | 492.20 | -13.92 | 7906 |
| Voc. & tech. students (mil) | 1230 | 111698.6 | 253483.79 | 0 | 2300769 |
| R&D expend. % of GDP | 1230 | .21 | .16 | .01 | .86 |
| R&D researchers (per mil) | 1230 | 162.65 | 225.9 | 5.94 | 1463.77 |
| R&D technicians (per mil) | 1230 | 57.02 | 63.01 | .13 | 627.73 |
| High-tech exports (mil) | 1230 | 6.23 | 9.29 | 0 | 68.14 |
| ECI (econ. complexity) | 1230 | -.72 | .63 | -3.04 | .82 |
| Tax revenue (% of GDP) | 1230 | 16.22 | 11.71 | 0 | 149.28 |
| Business startup cost | 1230 | 85.38 | 137.76 | 0 | 1314.6 |
| Domestic credit by banks | 1230 | 25.07 | 20.37 | .5 | 137.91 |
| Days to start business | 1230 | 35.34 | 37.71 | 1 | 260.5 |
| Days enforcing contract | 1230 | 666.61 | 329.52 | 225 | 1800 |
| Days to register property | 1230 | 87.33 | 97.58 | 1 | 690 |
| Openness measure | 1230 | .11 | .08 | .01 | .44 |
| Days to electric meter | 1230 | 37.24 | 33.64 | 2.5 | 194.3 |
| Business density | 1230 | 1.06 | 1.47 | .01 | 12.31 |
| Financial accountholders | 1230 | 30.94 | 22.53 | 1.52 | 92.97 |
| Commercial banks | 1230 | 10.49 | 11.99 | .27 | 71.23 |
| Primary enrollment (gross) | 1230 | 103.36 | 18.18 | 23.36 | 149.96 |
| Sec. enrollment (gross) | 1230 | 57.49 | 25.99 | 5.93 | 123.03 |
| Primary pupil-teacher ratio | 1230 | 34.43 | 14.36 | 8.68 | 100.24 |
| Primary completion rate | 1230 | 79.41 | 20.89 | 26.1 | 134.54 |
| Govt. expend. on educ. | 1230 | 4.36 | 2.22 | .69 | 12.9 |
| Human Capital Index 0-1 | 1230 | .42 | .09 | .29 | .69 |
| Advanced educ. labor | 1230 | 75.5 | 10.55 | 39.97 | 96.36 |
| Compulsory educ. (years) | 1230 | 8.45 | 2.16 | 4 | 15 |
| Industry employment | 1230 | 14.52 | 7 | .64 | 32.59 |
| Service employment | 1230 | 39.43 | 15.05 | 7.16 | 75.34 |
| Mobile subscriptions | 1230 | 59.12 | 38.15 | .26 | 181.33 |
| Access to electricity | 1230 | 57.02 | 31.3 | 1.24 | 100 |
| Broadband subscriptions | 1230 | 1.97 | 4.12 | 0 | 25.41 |
| Telephone subscriptions | 1230 | 5.31 | 7.39 | 0 | 32.85 |
| Energy use (per capita) | 1230 | 560.21 | 392.9 | 9.55 | 2246.92 |
| Logistic perf. Index 1-5 | 1230 | 2.18 | .33 | 1.1 | 3.34 |
| Internet users | 1230 | 16 | 16.3 | .03 | 89.44 |
| CPIA econ. mgmt. | 1230 | 3.39 | .69 | 1 | 5.5 |
| Public sect. mgmt. & instit | 1230 | 3.06 | .5 | 1.4 | 4.2 |
| Sructural policies | 1230 | 3.3 | .54 | 1.17 | 5 |
| Statistical capacity 0-100 | 1230 | 59.82 | 14.89 | 20 | 96.67 |
| Legal Rights Index 0-12 | 1230 | 4.83 | 3.1 | 0 | 11 |

| Variable | Obs | Mean | Std. Dev. | Min | Max |
|---|---|---|---|---|---|
| Human resources rating | 1230 | 3.52 | .63 | 1 | 4.5 |
| Equity of public resc use | 1230 | 3.38 | .64 | 1 | 4.5 |
| Social protection rating | 1230 | 3.03 | .59 | 1 | 4.5 |
| Social inclusion o.. | 1230 | 3.28 | .51 | 1.5 | 4.3 |
| National headcount poverty | 1230 | 38.52 | 15.13 | 4.1 | 82.3 |
| Social contributions | 1230 | 3.23 | 7.53 | 0 | 39.74 |
| GDP per capita 2010 | 1230 | 1969.31 | 1812.13 | 208.07 | 9350.75 |
| Log GDP per capita2010 | 1230 | 7.24 | .82 | 5.34 | 9.14 |
| Tech. coop. grants (mil) | 1230 | 92.57 | 108.20 | 0.51 | 1062 |
| Total population (mil) | 1230 | 34.61 | 141 | 0.01 | 1366 |
| Gross capital (mil) | 1230 | 15690 | 79490 | 0.00 | 991400 |
| Incoming tourists' no. (mil) | 1230 | 0.96 | 1.80 | 0.00 | 18.01 |
| Merch. imports frm HICs | 1230 | 49.31 | 20.07 | 2.5 | 99.56 |
| Net ODA/aid received (mil) | 1230 | 820.70 | 1058 | -247.40 | 11880 |
| Health expenditure | 1230 | 6.09 | 3.13 | 1.03 | 21.46 |
| No. of employers | 1230 | 2.24 | 2.31 | 0 | 13.76 |

**Appendix B. Tables B.1-B.6. Factor Loadings Tables from Six Confirmatory Analyses.**

The following six tables show factor loadings obtained after six confirmatory factor analyses (CFAs) that I conduct for six capacities. Loadings indicate correlations among one of the latent factors in the first row and the variables in the column in each table. Bold values indicate that these variables load highly on a particular factor in the first row of the table.

**Table B.1. Technology Capacity Factor Loadings**

| Variable | base sci & tech | medium sci & tech | high sci & tech |
|----------|-----------------|-------------------|-----------------|
| Sci & tech. articles | **0.9503** | 0.0479 | 0.0704 |
| Intellectual payments (mil) | **0.933** | 0.0362 | 0.051 |
| Voc. & tech. students (mil) | **0.7254** | 0.0288 | -0.0843 |
| R&D expend. % of GDP | 0.4214 | 0.0703 | **0.5683** |
| R&D researchers (per mil) | -0.0316 | **0.8274** | -0.0003 |
| R&D technicians (per mil) | 0.1027 | 0.2936 | **0.5956** |
| High-tech exports (mil) | -0.0749 | -0.0485 | **0.7651** |
| ECI (econ. complexity) | 0.1288 | **0.8233** | 0.0789 |

**Table B.2. Human Capital Capacity Factor Loadings**

| Variable | Specialized skills | Generalized skills |
|---|---|---|
| Primary enrollment (gross) | 0.2055 | **0.8694** |
| Sec. enrollment (gross) | **0.9331** | 0.0169 |
| Primary pupil-teacher ratio | -0.815 | **0.1271** |
| Primary completion rate | **0.8724** | 0.2845 |
| Govt. expend. on educ. | **0.417** | 0.07 |
| Human Capital Index 0-1 | **0.8332** | 0.0821 |
| Advanced educ. labor | -0.1977 | **0.1646** |
| Compulsory educ. (years) | **0.3599** | -0.6409 |
| Industry employment | **0.7147** | -0.2713 |
| Service employment | **0.7156** | -0.3428 |

**Table B.3. Public Policy Capacity Factor Loadings**

| Variable | Public policy (inc. fiscal, monetary, structural policies…) |
|---|---|
| Statistical capacity 0-100 | **0.7359** |
| CPIA econ. mgmt. | **0.8166** |
| Public sect. mgmt.& instit. | **0.8591** |
| Structural policies | **0.8923** |
| Legal Rights Index 0-12 | **0.4181** |

**Table B.4. Social Policy Capacity Factor Loadings**

| Variable | Social capacity (inc. equity, inclusion…) |
|---|---|
| Human resources rating | **0.8751** |
| Equity of public resc use | **0.8328** |
| Social protection rating | **0.8605** |
| Social inclusion (inc. gender equity and others) | **0.9724** |
| National headcount poverty | **-0.48** |
| Social contributions | **0.3669** |

**Table B.5. Infrastructural Capacity Factor Loadings**

| Variable | Infrastructure (ICT & energy) | Logistic Perf. Index (trade & transport infras.) |
|---|---|---|
| Mobile subscriptions | **0.5603** | 0.5062 |
| Access to electricity | **0.7411** | 0.3341 |
| Broadband subscriptions | **0.8804** | 0.024 |
| Telephone subscriptions | **0.8376** | -0.0125 |
| Energy use (per capita) | **0.8229** | 0.0057 |
| Logistic perf. Index 1-5 | 0.009 | **0.9091** |
| Internet users | **0.8129** | 0.3437 |

**Tabel B.6. Financial Capacity Factor Loadings**

| Variable | Financial infrastructure | Financial environment | Strength of financial regulation | Financial enabling environment |
|---|---|---|---|---|
| Tax revenue (% of GDP) | 0.3541 | 0.2798 | **0.5426** | -0.4327 |
| Business startup cost | -0.4405 | **0.5017** | 0.0394 | -0.0492 |
| Domestic credit by banks | **0.7692** | 0.1594 | -0.1514 | 0.1406 |
| Days to start business | -0.1497 | **0.8564** | 0.2287 | -0.1117 |
| Days enforcing contract | -0.095 | -0.0815 | **0.8511** | 0.1189 |
| Days to register property | -0.2421 | 0.3618 | **0.3821** | 0.3226 |
| Openness measure | 0.3364 | **0.7956** | -0.3945 | -0.0183 |
| Days to electric meter | 0.1188 | -0.0908 | 0.0847 | **0.8499** |
| Business density | **0.6787** | -0.0578 | 0.0196 | -0.224 |
| Financial accountholders | **0.765** | -0.0357 | 0.1769 | 0.0729 |
| Commercial banks | **0.7567** | -0.034 | -0.2106 | 0.0268 |

**Figure C.1. Scree Plot Obtained via CFA for Technology Capacity- This plot suggests retaining three factors.**

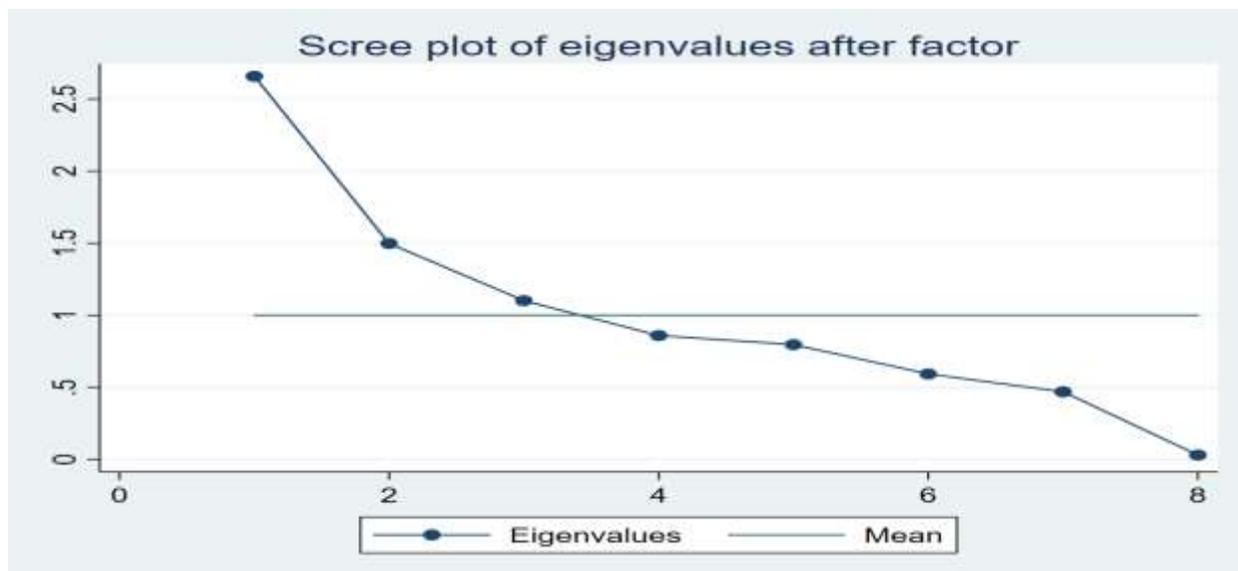

**Figure C.2. Scree Plot Obtained via CFA for Human Capital Capacity- This plot suggests retaining two factors.**

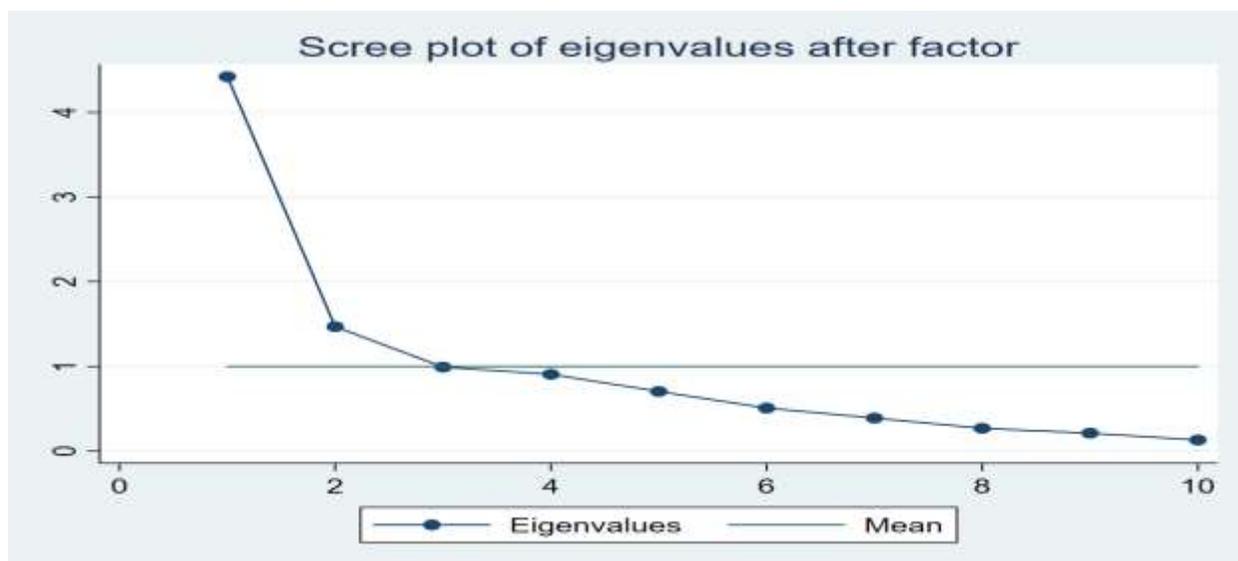

---

[24] Scree Plot is a powerful visual tool for determining the number of factors to be retained. It is basically a plot of the eigenvalues shown in decreasing order. The eigenvalue is calculated by summing the squared factor loadings for that factor. Factor loadings, on the other hand, are basically correlations among the variables and their latent constructed factors. Factors having eigenvalues greater than 1 are retained.

**Figure C.3. Scree Plot Obtained via CFA for Infrastructural Capacity- This plot suggests retaining two factors**

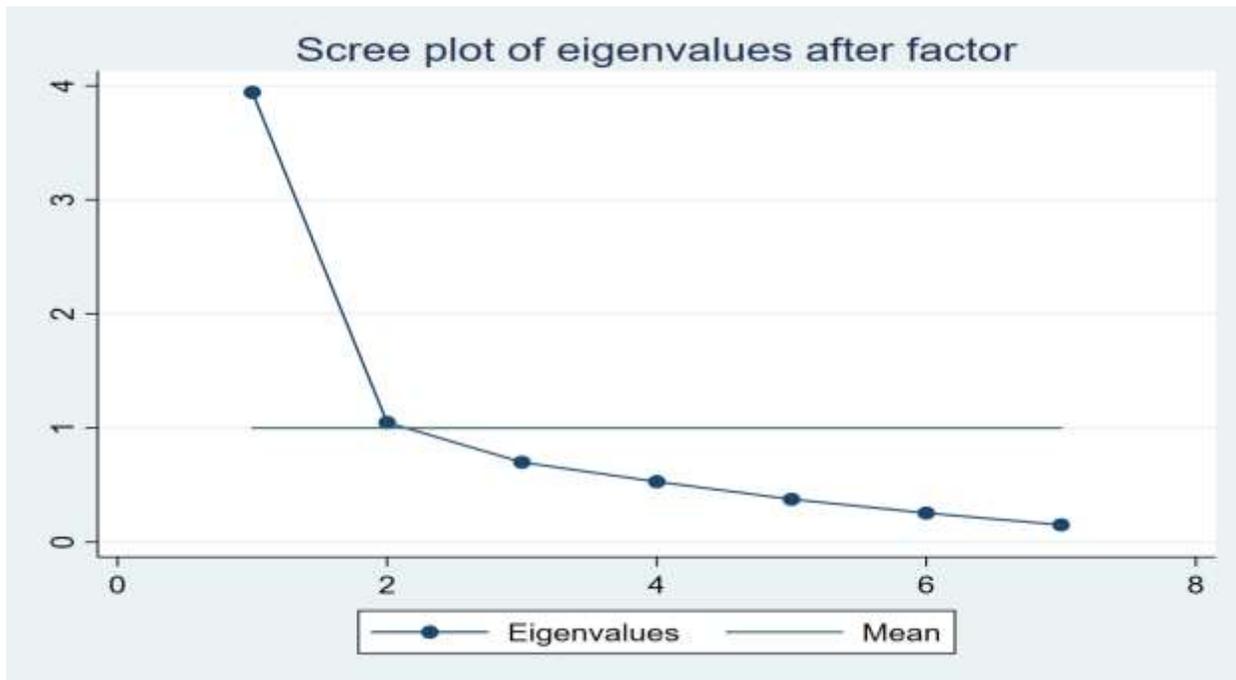

**Figure C.4. Scree Plot Obtained via CFA for Financial Capacity-This plot suggests retaining four factors**

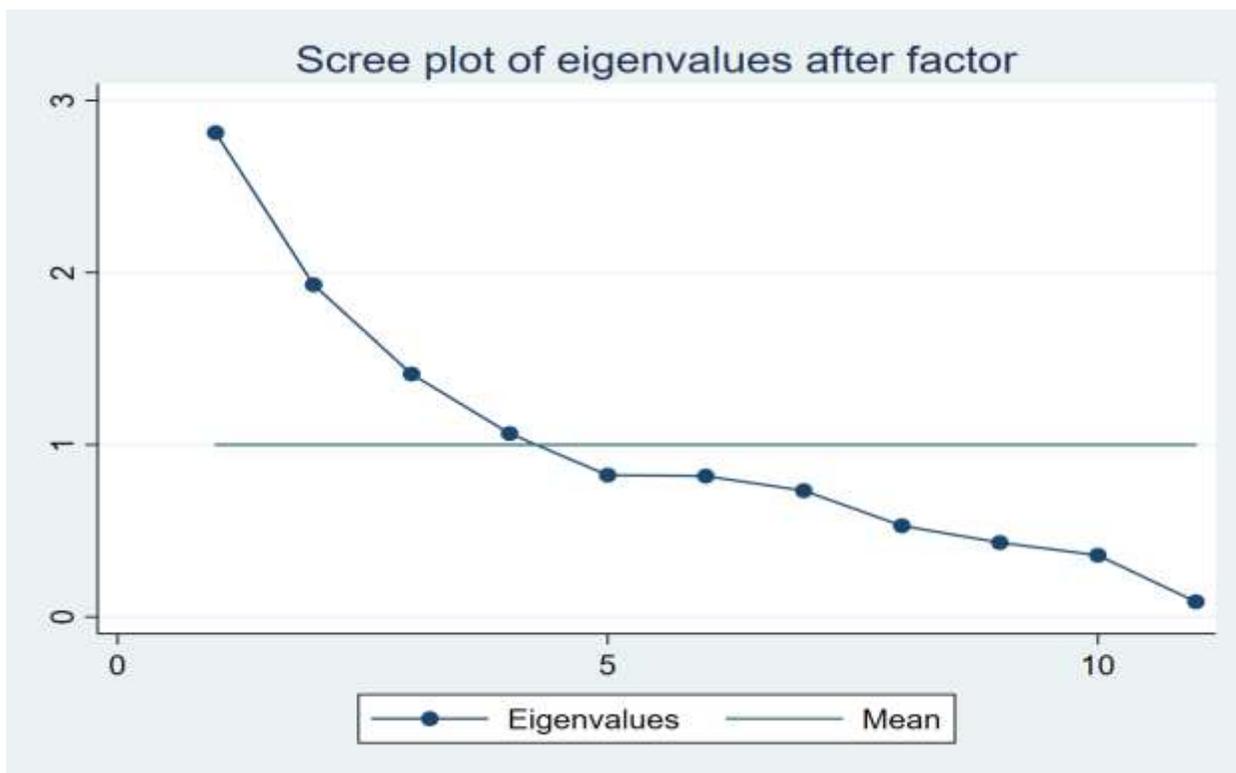

**Figure C.5. Scree Plot Obtained via CFA for Public Policy Capacity- this plot suggests retaining one factor**

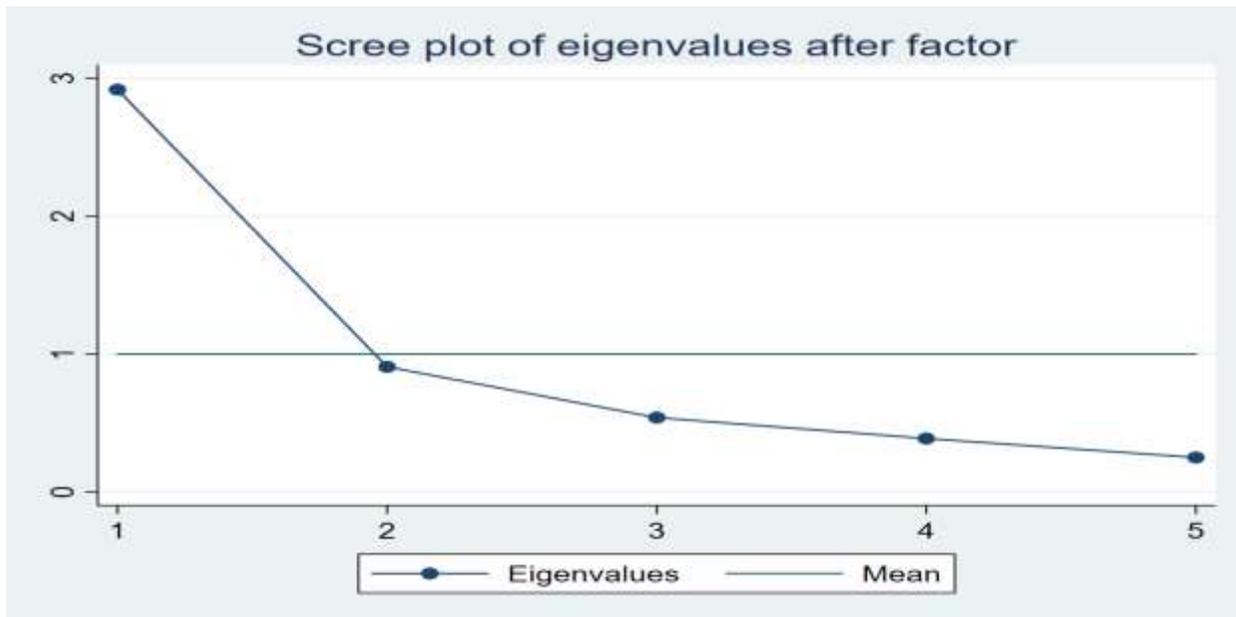

**Figure C.6. Scree Plot Obtained via CFA for Social Capacity- This plot suggests retaining one factor**

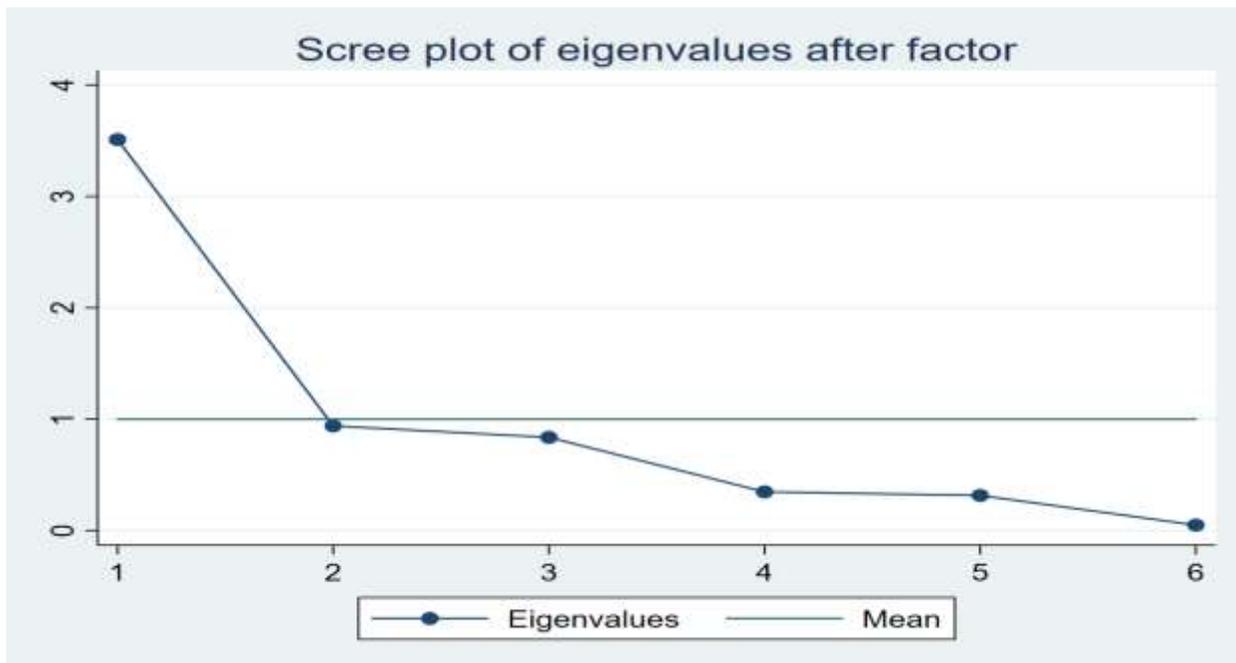

**Appendix D. Table Main Results: Full-scale Table. Dependent Variable, Log of GDP Per Capita.**

| VARIABLES | Pooled OLS | Random Effects | Fixed Effects |
|---|---|---|---|
| | | | |
| Public policy (inc. fiscal, monetary, structural, etc.) | 0.098*** | 0.077*** | 0.087*** |
| | (0.022) | (0.025) | (0.027) |
| Infrastructure (ICT & energy) | 0.371*** | 0.134*** | 0.095*** |
| | (0.026) | (0.024) | (0.028) |
| Logistic Per. Index (trade & transp. infras.) | 0.100*** | 0.037*** | 0.029*** |
| | (0.016) | (0.009) | (0.009) |
| Specialized skills | 0.240*** | 0.111*** | 0.061*** |
| | (0.024) | (0.018) | (0.018) |
| Generalized skills | -0.081*** | -0.030** | -0.018 |
| | (0.012) | (0.015) | (0.014) |
| Financial infrastructure | 0.109*** | 0.037*** | 0.025** |
| | (0.017) | (0.013) | (0.012) |
| Financial environment | 0.047*** | 0.023** | 0.025** |
| | (0.015) | (0.010) | (0.010) |
| Strength of financial regulations | -0.045*** | 0.010 | 0.010 |
| | (0.014) | (0.018) | (0.018) |
| Enabling financial environment | 0.036*** | 0.003 | 0.002 |
| | (0.011) | (0.005) | (0.005) |
| Base sci & tech | -0.179*** | -0.023 | -0.028 |
| | (0.034) | (0.023) | (0.025) |
| Medium sci & tech | -0.127*** | -0.001 | 0.002 |
| | (0.017) | (0.013) | (0.013) |
| High sci & tech | -0.061*** | -0.002 | 0.001 |
| | (0.014) | (0.006) | (0.006) |
| Social capacity (incl. equity, inclusion, etc.) | -0.128*** | 0.004 | -0.001 |
| | (0.023) | (0.019) | (0.018) |
| Tech. coop. grants (st) | 0.006 | -0.024 | -0.017 |
| | (0.021) | (0.017) | (0.018) |
| Total population (st) | -0.373*** | -0.143*** | 0.388 |
| | (0.058) | (0.046) | (0.345) |
| Gross capital (st) | 0.569*** | 0.155*** | 0.035 |
| | (0.064) | (0.054) | (0.063) |
| Incoming tourists' no. (st) | -0.049*** | 0.005 | 0.021 |
| | (0.018) | (0.012) | (0.013) |
| Merch. imports frm HICs (st) | 0.170*** | 0.078*** | 0.058*** |
| | (0.014) | (0.021) | (0.022) |
| Net ODA/aid received (st) | -0.028 | -0.005 | -0.001 |
| | (0.030) | (0.017) | (0.019) |
| Health expenditure (st) | -0.086*** | -0.054*** | -0.050*** |
| | (0.012) | (0.015) | (0.016) |

| VARIABLES | Pooled OLS | Random Effects | Fixed Effects |
|---|---|---|---|
| No. of employers (st) | 0.027** | 0.037*** | 0.029*** |
|  | (0.012) | (0.010) | (0.010) |
| YR2006 | -0.020 | 0.005 | 0.011 |
|  | (0.065) | (0.014) | (0.014) |
| YR2007 | -0.037 | 0.015 | 0.033* |
|  | (0.065) | (0.017) | (0.017) |
| YR2008 | -0.043 | 0.033** | 0.053*** |
|  | (0.064) | (0.017) | (0.017) |
| YR2009 | -0.070 | 0.020 | 0.040* |
|  | (0.064) | (0.021) | (0.022) |
| YR2010 | -0.052 | 0.046* | 0.071*** |
|  | (0.065) | (0.024) | (0.023) |
| YR2011 | -0.104 | 0.047* | 0.080*** |
|  | (0.063) | (0.025) | (0.026) |
| YR2012 | -0.109* | 0.064** | 0.100*** |
|  | (0.063) | (0.031) | (0.034) |
| YR2013 | -0.077 | 0.094** | 0.132*** |
|  | (0.066) | (0.039) | (0.041) |
| YR2014 | -0.103 | 0.092** | 0.134*** |
|  | (0.066) | (0.040) | (0.042) |
| YR2015 | -0.129* | 0.082* | 0.128** |
|  | (0.068) | (0.045) | (0.050) |
| YR2016 | -0.120* | 0.090** | 0.135** |
|  | (0.067) | (0.045) | (0.052) |
| YR2017 | -0.167** | 0.102** | 0.155*** |
|  | (0.069) | (0.050) | (0.056) |
| YR2018 | -0.171** | 0.093* | 0.147** |
|  | (0.071) | (0.050) | (0.057) |
| YR2019 | -0.115 | 0.115** | 0.166** |
|  | (0.072) | (0.058) | (0.065) |
| Constant | 7.329*** | 7.181*** | 7.149*** |
|  | (0.051) | (0.068) | (0.032) |
|  |  |  |  |
| Observations | 1,230 | 1,230 | 1,230 |
| R-squared | 0.799 | 0.727 | 0.468 |
| Controls | YES | YES | YES |
| Year Effects | YES | YES | YES |
| Country Fixed Effects | NO | NO | YES |
| Number of countries | 82 | 82 | 82 |

Robust errors in parentheses
*** p<0.01, ** p<0.05, * p<0.1

**Appendix E. Table. Detailed Descriptive Statistics of Five Clusters Within LMICs.**

| CAPACITY FACTORS AND LMICs CLUSTERS | N | Mean | SD | Min | Max |
|---|---|---|---|---|---|
| **1. LEADING ECONOMIES** | | | | | |
| Public Policy Factor (incl. fiscal, monetary, structural policies…) | 11 | .85 | .43 | .37 | 1.51 |
| Infrastructure Capacity Factor (infras. inc. ICT & energy…) | 11 | 2.65 | .72 | 1.48 | 3.62 |
| Human Capacity Factor (specialized skills inc. industry, service employment…) | 11 | 1.45 | .3 | 1.06 | 1.89 |
| Financial Capacity Factor (financial infras. inc. banks, credit, businesses…) | 11 | 1.67 | .84 | .73 | 3.05 |
| **2. WALKING ECONOMIES** | | | | | |
| Public Policy Factor (incl. fiscal, monetary, structural policies…) | 13 | .59 | .41 | -.34 | 1.21 |
| Infrastructure Capacity Factor (infras. inc. ICT & energy…) | 13 | .73 | .56 | -.07 | 1.89 |
| Human Capacity Factor (specialized skills inc. industry, service employment…) | 13 | 1.03 | .33 | .47 | 1.32 |
| Financial Capacity Factor (financial infras. inc. banks, credit, businesses…) | 13 | 1.12 | .55 | .08 | 2.23 |
| **3. CREEPING ECONOMIES** | | | | | |
| Public Policy Factor (incl. fiscal, monetary, structural policies…) | 12 | -.43 | .47 | -1.28 | .37 |
| Infrastructure Capacity Factor (infras. inc. ICT & energy…) | 12 | .33 | .53 | -.25 | 1.4 |
| Human Capacity Factor (specialized skills inc. industry, service employment…) | 12 | .72 | .35 | .16 | 1.17 |
| Financial Capacity Factor (financial infras. inc. banks, credit, businesses…) | 12 | .05 | .51 | -.6 | 1.1 |
| **4. CRAWLING ECONOMIES** | | | | | |
| Public Policy Factor (incl. fiscal, monetary, structural policies…) | 30 | .25 | .49 | -.5 | 1.57 |
| Infrastructure Capacity Factor (infras. inc. ICT & energy…) | 30 | -.4 | .29 | -.92 | .21 |
| Human Capacity Factor (specialized skills inc. industry, service employment…) | 30 | -.42 | .49 | -1.23 | .66 |
| Financial Capacity Factor (financial infras. inc. banks, credit, businesses…) | 30 | -.22 | .38 | -.85 | .91 |
| **5 SLEEPING ECONOMIES** | | | | | |
| Public Policy Factor (incl. fiscal, monetary, structural policies…) | 16 | -1.44 | .81 | -3.07 | -.73 |
| Infrastructure Capacity Factor (infras. inc. ICT & energy…) | 16 | -.45 | .3 | -.89 | .07 |
| Human Capacity Factor (specialized skills inc. industry, service employment…) | 16 | -.62 | .56 | -1.68 | .33 |
| Financial Capacity Factor (financial infras. inc. banks, credit, businesses…) | 16 | -.66 | .32 | -1.06 | -.01 |

## Appendix F. Figure F.1. and Table F.1.

The Elbow Rule suggests that 5 clusters solution is an optimal solution. Figure F.1. is produced at seed 1011. This figure shows at k=5, a large drop occurs in the Within Sum of Square (Intra Cluster Variation)

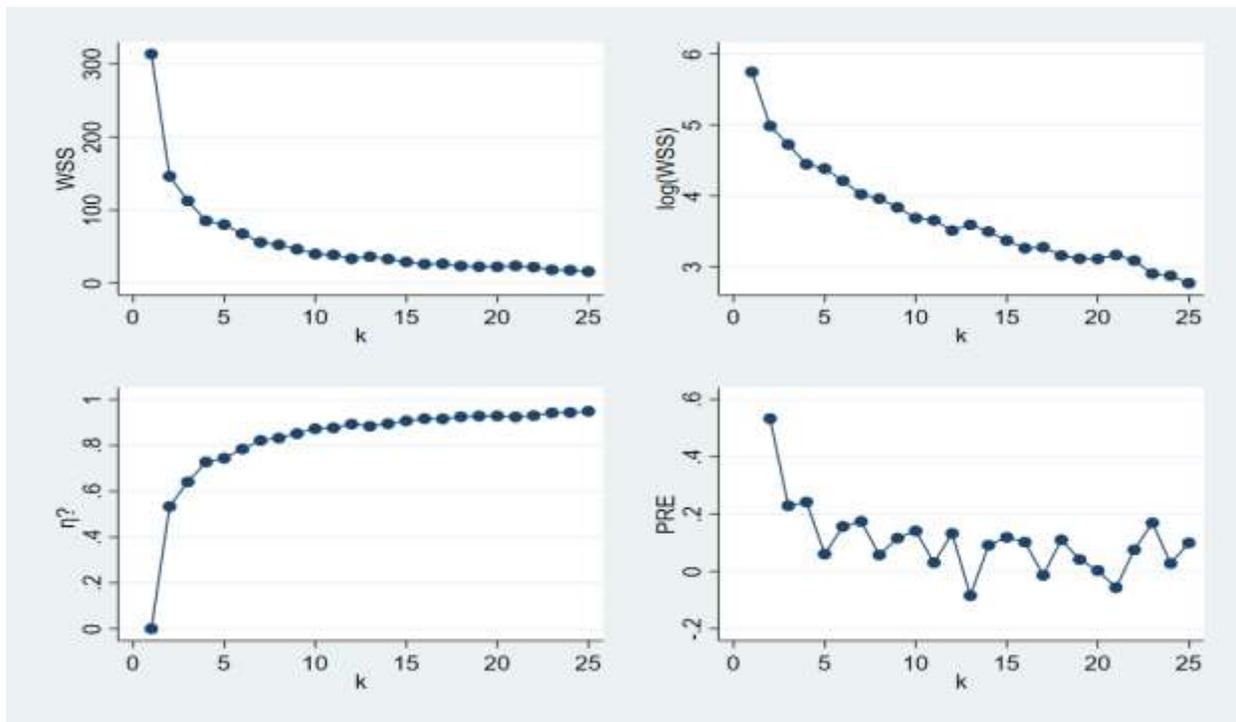

**Figure F.1.** WSS, log(WSS), η2, and PRE for all K cluster solutions (seed 1011)

## Table F.1. Corresponding matrix list with Seed 1011.

The table shows the highest drop in WSS at k=5. To learn more about all the diagnostics in the table, refer to (Makles 2012).

| WSS[25,5] | | | | | |
|---|---|---|---|---|---|
| | k | WSS | log(WSS) | eta-squared | PRE |
| r1 | 1 | 313.51995 | 5.747863 | 0 | . |
| r2 | 2 | 146.26096 | 4.9853924 | .53348756 | .53348756 |
| r3 | 3 | 112.68659 | 4.7246104 | .64057602 | .22955114 |
| r4 | 4 | 85.396229 | 4.4473019 | .72762107 | .24217929 |
| r5 | 5 | 80.222239 | 4.3848008 | .74412397 | .06058804 |
| r6 | 6 | 67.619651 | 4.2138986 | .78432106 | .15709594 |
| r7 | 7 | 55.816877 | 4.0220763 | .82196706 | .17454651 |
| r8 | 8 | 52.592082 | 3.9625656 | .83225284 | .05777455 |

Output terminated

## Appendix F. Figure F.2. and Table F.2

The Elbow Rule suggests that 5 clusters solution is an optimal solution. Figure F.2. is produced at seed 789. This figure shows at k=5, a large drop occurs in the Within Sum of Square (Intra Cluster Variation)

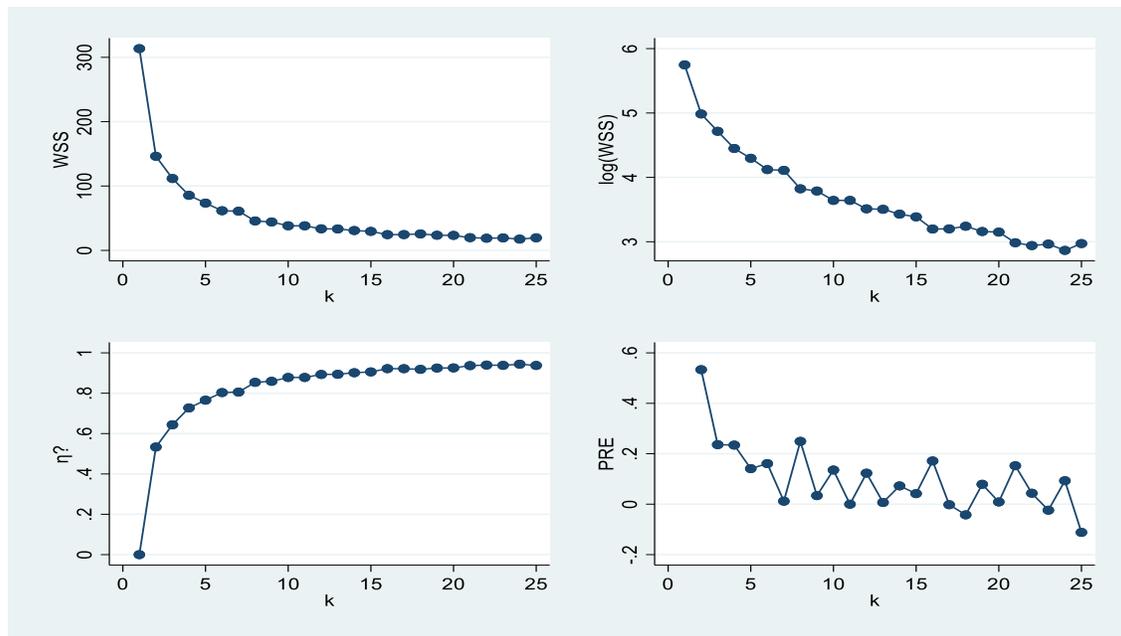

**Figure F.2.** WSS, log(WSS), η2, and PRE for all K cluster solutions (generated with random number 789)

### Table F.2. Corresponding matrix list with Seed 789.

The table shows the highest drop in WSS at k=5. To learn more about all the diagnostics in the table, refer to (Makles 2012).

| WSS[25,5] | | | | | |
|---|---|---|---|---|---|
| | k | WSS | log(WSS) | eta-squared | PRE |
| r1 | 1 | 313.52 | 5.747863 | 0 | . |
| r2 | 2 | 146.261 | 4.985392 | 0.533488 | 0.533488 |
| r3 | 3 | 111.7259 | 4.716048 | 0.64364 | 0.23612 |
| r4 | 4 | 85.541 | 4.448996 | 0.727159 | 0.234367 |
| r5 | 5 | 73.47712 | 4.296974 | 0.765638 | 0.14103 |
| r6 | 6 | 61.65221 | 4.121509 | 0.803355 | 0.160933 |
| r7 | 7 | 60.91858 | 4.109538 | 0.805695 | 0.011899 |
| r8 | 8 | 45.7367 | 3.822901 | 0.854119 | 0.249216 |
| r9 | 9 | 44.18006 | 3.788274 | 0.859084 | 0.034035 |

Output terminated

**Appendix G. Table. Exploratory Factor Analysis Returned 12 Factors.[25]**

| Variable | specialized skills plus financial infras. | Public policy and social capacities | Overall infrastructu-re(1) | Base science & tech | Factor5 | Factor6 | Overall infrastruct-ure(2) | Legal rights strength | High sci & tech and infras. | Factor 10 | Factor 11 | Factor 12 |
|---|---|---|---|---|---|---|---|---|---|---|---|---|
| Sci & tech. articles | 0.0207 | 0.0764 | 0.0517 | **0.9397** | 0.02 | -0.0459 | 0.004 | 0.039 | 0.0585 | 0.033 | 0.0335 | -0.0467 |
| Intellectual payments (mil) | 0.0329 | 0.0566 | 0.0536 | **0.9234** | 0.0514 | -0.0352 | 0.0245 | 0.0537 | 0.03 | 0.0387 | -0.0158 | -0.0459 |
| Voc. & tech. students (mil) | 0.005 | 0.0832 | 0.2903 | **0.6875** | 0.1596 | 0.0503 | -0.0821 | -0.1692 | -0.1118 | -0.139 | 0.1024 | -0.0174 |
| R&D expend. % GDP | 0.2173 | 0.1685 | 0.1221 | 0.3663 | 0.2231 | 0.0504 | 0.0388 | 0.068 | 0.1569 | -0.1429 | **0.4843** | -0.3093 |
| R&D researchers (per mil) | 0.328 | 0.0992 | **0.5451** | 0.0015 | 0.1015 | 0.1249 | 0.0934 | 0.3916 | 0.1507 | -0.0245 | -0.0859 | -0.1545 |
| R&D technicians (per mil) | 0.0908 | 0.0007 | 0.0825 | 0.1298 | 0.1346 | -0.0225 | -0.0138 | 0.0363 | **0.7312** | -0.1723 | 0.2262 | -0.0566 |
| High-tech exports (mil) | 0.0855 | 0.0856 | 0.0007 | 0.0235 | **0.8081** | 0.1657 | -0.0219 | -0.0469 | 0.0516 | -0.0493 | 0.0302 | -0.0302 |
| ECI (econ. complexity) | 0.3267 | 0.2288 | 0.2594 | 0.1837 | 0.1177 | -0.0963 | -0.2111 | 0.385 | **0.3713** | 0.0908 | -0.1736 | 0.2039 |
| Tax revenue (% of GDP) | 0.1016 | 0.0307 | 0.0013 | 0.0021 | **0.8555** | 0.0499 | -0.0437 | 0.0359 | 0.0643 | 0.0905 | -0.1068 | 0.0597 |
| Business startup cost | 0.3716 | 0.2756 | 0.0372 | 0.1592 | 0.1096 | 0.2865 | -0.1796 | -0.3803 | 0.3488 | -0.0859 | -0.124 | 0.0816 |
| Domestic credit by banks | **0.5465** | 0.326 | 0.0658 | 0.1413 | 0.101 | 0.3119 | 0.1963 | 0.2433 | 0.1902 | 0.0579 | 0.0298 | 0.0447 |
| Days to start business | 0.1513 | 0.2932 | 0.0005 | 0.0772 | 0.3535 | **0.6143** | -0.1808 | -0.3413 | 0.0105 | 0.0705 | 0.057 | 0.1574 |
| Days enforcing contract | 0.0294 | 0.2718 | 0.1481 | 0.3224 | 0.3195 | **-0.4383** | -0.0012 | -0.1728 | 0.0914 | 0.1731 | 0.1768 | 0.2891 |
| Days to register property | 0.0391 | 0.2637 | 0.1273 | 0.0187 | 0.048 | 0.0284 | -0.7394 | -0.0157 | 0.0061 | 0.1617 | 0.1503 | 0.0742 |
| Openness measure | 0.1957 | 0.0423 | 0.0647 | 0.0714 | 0.087 | **0.9246** | 0.0289 | 0.0279 | -0.0014 | 0.0635 | -0.0348 | 0.0541 |
| Days to electric meter | 0.0215 | 0.1558 | 0.0188 | 0.0356 | 0.1806 | -0.0752 | -0.0645 | 0.0733 | 0.1203 | 0.1072 | **0.7506** | 0.0985 |
| Business density | 0.3393 | 0.0729 | 0.2593 | 0.1768 | 0.2736 | 0.0021 | 0.3118 | 0.0602 | 0.422 | 0.114 | 0.0838 | -0.318 |
| Financial accountholders | **0.5086** | 0.0983 | 0.257 | 0.1135 | 0.2207 | -0.0332 | 0.2937 | -0.0833 | -0.0063 | 0.0674 | 0.2658 | 0.1544 |
| Commercial banks | **0.5589** | 0.1767 | 0.657 | 0.0005 | 0.0077 | 0.0632 | 0.0569 | 0.0426 | -0.0158 | -0.0476 | -0.0121 | -0.0874 |
| Primary enrollment (gross) | 0.0508 | 0.1639 | 0.0299 | 0.0242 | 0.0639 | 0.0674 | -0.0985 | 0.0561 | -0.0499 | **0.8715** | 0.0324 | 0.0169 |
| Sec. enrollment (gross) | **0.8638** | 0.1875 | 0.27 | 0.0234 | 0.0282 | 0.0014 | -0.0406 | 0.0044 | 0.023 | 0.1212 | -0.0491 | -0.024 |

[25] I do not name all the factors for space constraints. I generally name those factors which correspond to factors from confirmatory factor analyses. Factors loadings indicate correlations between latent factors in the first row and variables in the first column. The bold values indicate high values of correlation. Based on these values, the latent factor is named accordingly to represent the variables.

| Variable | specialized skills plus financial infras. | Public policy and social capacities | Overall infrastructu-re(1) | Base science & tech | Factor5 | Factor6 | Overall infrastruct-ure(2) | Legal rights strength | High sci & tech and infras. | Factor 10 | Factor 11 | Factor 12 |
|---|---|---|---|---|---|---|---|---|---|---|---|---|
| Primary pupil-teacher ratio | **0.8053** | 0.1067 | 0.0438 | 0.0308 | 0.109 | -0.0176 | 0.0864 | 0.0597 | -0.0444 | 0.0586 | 0.0797 | 0.1484 |
| Primary completion rate | **0.7638** | 0.1652 | 0.1194 | 0.0598 | 0.0067 | 0.1846 | -0.1491 | 0.1555 | -0.0848 | 0.3245 | 0.0115 | -0.0832 |
| Govt. expend. on educ. | 0.28 | 0.0695 | 0.2493 | 0.0028 | 0.3787 | -0.0429 | **-0.5468** | 0.1275 | -0.1532 | 0.0295 | -0.1306 | -0.1087 |
| Human Capital Index 0-1 | **0.7566** | 0.2144 | 0.2217 | 0.0518 | 0.0144 | 0.2599 | 0.121 | 0.1856 | 0.0118 | 0.1465 | -0.0874 | 0.0446 |
| Advanced educ. labor | 0.1669 | 0.1144 | 0.0384 | 0.1449 | 0.0477 | 0.1157 | -0.0041 | 0.0347 | -0.018 | -0.0041 | 0.036 | **0.8409** |
| Compulsory educ. (years) | 0.296 | 0.1469 | 0.307 | 0.0351 | 0.0189 | -0.0217 | -0.0197 | 0.3526 | 0.0905 | -0.5659 | -0.0281 | 0.0414 |
| Industry employment | **0.7342** | 0.1636 | 0.0579 | 0.1594 | 0.076 | -0.0566 | 0.0121 | -0.1152 | -0.0131 | -0.1765 | 0.089 | -0.0011 |
| Service employment | **0.7784** | 0.0005 | 0.0051 | 0.1305 | 0.0378 | -0.1441 | -0.1718 | -0.1265 | 0.1481 | -0.2116 | 0.1429 | -0.0938 |
| Mobile subscriptions | **0.5763** | 0.2221 | 0.0222 | 0.0247 | 0.0267 | -0.0502 | 0.4972 | 0.1907 | -0.0375 | -0.0176 | 0.0887 | 0.0305 |
| Statistical capacity 0-100 | 0.0353 | **0.6782** | 0.1537 | 0.1135 | 0.0117 | -0.0401 | 0.3585 | 0.0849 | -0.1397 | 0.0044 | 0.0338 | -0.0091 |
| Access to electricity | **0.8654** | 0.2106 | 0.1639 | 0.0788 | 0.0229 | 0.0366 | -0.008 | 0.088 | -0.046 | -0.0924 | -0.014 | -0.0517 |
| Broadbamd subscriptions | **0.5305** | 0.1346 | 0.3801 | 0.0513 | 0.0107 | 0.0425 | 0.1857 | 0.1755 | 0.4828 | 0.0226 | -0.0711 | 0.0774 |
| Telephone subscriptions | **0.6531** | 0.2577 | 0.3271 | 0.0964 | 0.0411 | 0.0434 | 0.012 | 0.0131 | 0.3314 | -0.0094 | -0.1537 | 0.0028 |
| Energy use | 0.4826 | 0.0926 | **0.6948** | 0.0226 | 0.0753 | 0.056 | 0.0694 | 0.159 | 0.2115 | 0.0617 | 0.1077 | -0.0202 |
| Logistic perf. Index 1-5 | 0.1844 | 0.2732 | 0.0443 | 0.2637 | 0.0407 | -0.0387 | -0.0249 | 0.3029 | -0.179 | -0.1601 | 0.3084 | 0.0029 |
| Internet users | **0.6748** | 0.1579 | 0.2608 | 0.0186 | 0.0676 | -0.0349 | 0.2516 | 0.1875 | 0.1365 | -0.0852 | 0.022 | 0.0083 |
| CPIA econ. mgmt. | 0.0378 | **0.8201** | 0.0014 | 0.0857 | 0.0099 | 0.0206 | -0.0848 | 0.0246 | -0.0413 | -0.0401 | -0.0413 | -0.0255 |
| Public sect. mgmt. & instit | 0.3208 | **0.8169** | 0.0329 | 0.0108 | 0.0365 | -0.0148 | -0.028 | 0.1423 | 0.136 | 0.0115 | 0.0328 | 0.0339 |
| Structural policies | 0.2167 | **0.7876** | 0.0149 | 0.143 | 0.0023 | -0.0597 | 0.0427 | 0.0767 | 0.2125 | -0.0359 | 0.1076 | -0.0057 |
| Legal Rights Index 0-12 | 0.0344 | **0.1705** | 0.0886 | 0.0418 | 0.0543 | -0.019 | 0.0045 | 0.7368 | 0.0773 | -0.0011 | 0.125 | 0.0483 |
| Human resources rating | 0.3217 | **0.7402** | 0.0807 | 0.0924 | 0.0582 | -0.0466 | 0.2279 | 0.0154 | 0.0583 | 0.1122 | -0.0252 | 0.0445 |
| Equity of public resc use | 0.0769 | **0.8575** | 0.0149 | 0.1193 | 0.0254 | 0.0661 | 0.0035 | 0.0786 | -0.1037 | 0.0571 | 0.0064 | 0.0451 |
| Social protection rating | 0.1675 | **0.8214** | 0.1794 | 0.0012 | 0.0914 | -0.0318 | -0.0101 | -0.0613 | 0.007 | -0.0348 | 0.1504 | -0.017 |
| Social inclusion | 0.2522 | **0.9065** | 0.1324 | 0.0621 | 0.0204 | 0.0495 | 0.0975 | 0.0526 | -0.0139 | 0.0981 | 0.0036 | 0.0399 |
| National headcount poverty | **0.541** | 0.2239 | 0.0882 | 0.0619 | 0.0164 | -0.1677 | -0.1796 | -0.0981 | 0.0002 | 0.082 | -0.1755 | -0.1342 |
| Social contributions | 0.3156 | 0.1568 | 0.7707 | 0.0253 | 0.0372 | 0.0386 | -0.0585 | -0.0897 | -0.1406 | -0.0584 | 0.0198 | 0.0164 |

**Appendix H. Table. Sensitivity Analysis Regressions with Robust Errors Including Factors from Exploratory Factor Analysis. Dependent Variable, Log of GDP Per Capita.**[26]

| VARIABLES | Pooled OLS | Random Effects | Random Effects |
|---|---|---|---|
| specialized skills plus financial infras. | 0.523*** | 0.236*** | 0.149*** |
| | (0.015) | (0.027) | (0.032) |
| **Public policy and social capacities** | 0.076*** | 0.115*** | 0.098*** |
| | (0.012) | (0.025) | (0.031) |
| **Overall infrastructure** | 0.154*** | 0.078*** | 0.050*** |
| | (0.010) | (0.014) | (0.014) |
| Base science & tech | -0.290*** | 0.014 | 0.025 |
| | (0.034) | (0.022) | (0.026) |
| Scores for factor 5 | -0.037*** | 0.017 | 0.017 |
| | (0.009) | (0.010) | (0.010) |
| Scores for factor 6 | 0.088*** | 0.022 | 0.020 |
| | (0.014) | (0.014) | (0.015) |
| Overall infrastructure | 0.024* | 0.020* | 0.021* |
| | (0.015) | (0.012) | (0.012) |
| Legal rights strength | 0.051*** | 0.042*** | 0.036*** |
| | (0.014) | (0.009) | (0.010) |
| High science & tech and infras. | 0.057*** | 0.044*** | 0.033*** |
| | (0.011) | (0.009) | (0.012) |
| Scores for factor 10 | -0.059*** | -0.008 | -0.000 |
| | (0.011) | (0.015) | (0.015) |
| Scores for factor 11 | 0.030** | 0.008 | 0.008 |
| | (0.012) | (0.007) | (0.007) |
| Scores for factor 12 | -0.053*** | 0.000 | 0.005 |
| | (0.011) | (0.008) | (0.008) |
| | (0.012) | (0.010) | (0.010) |
| Constant | 7.266*** | 7.169*** | 7.140*** |
| | (0.050) | (0.067) | (0.027) |
| | | | |
| Observations | 1,230 | 1,230 | 1,230 |
| R-squared | 0.793 | 0.706 | 0.448 |
| Controls | YES | YES | YES |
| Year Fixed Effects | YES | YES | YES |
| Country Fixed Effects | NO | NO | YES |
| Errors | Robust | Robust | Robust |
| Number of countries | 82 | 82 | 82 |

Standard errors in parentheses
*** p<0.01, ** p<0.05, * p<0.1

---

[26] I do not name all the factors. I generally name those factors which are significant and correspond to factors from confirmatory factor analyses.

**Appendix I. Table. Sensitivity Analysis: Regressions with Standard Errors. Dependent Variable, Log of GDP Per Capita.**

| VARIABLES | Pooled OLS | Random Effects | Fixed Effects |
|---|---|---|---|
| Public policy (inc. fiscal, monetary, structural…) | 0.098*** | 0.077*** | 0.087*** |
| | (0.021) | (0.013) | (0.013) |
| Infrastructure (ICT & energy) | 0.371*** | 0.134*** | 0.095*** |
| | (0.026) | (0.017) | (0.017) |
| Logistic Per. Index (trade & transp. infras.) | 0.100*** | 0.037*** | 0.029*** |
| | (0.016) | (0.007) | (0.007) |
| Specialized skills | 0.240*** | 0.111*** | 0.061*** |
| | (0.022) | (0.016) | (0.016) |
| Generalized skills | -0.081*** | -0.030*** | -0.018** |
| | (0.013) | (0.009) | (0.009) |
| Financial infrastructure | 0.109*** | 0.037*** | 0.025** |
| | (0.019) | (0.011) | (0.011) |
| Financial environment | 0.047*** | 0.023*** | 0.025*** |
| | (0.013) | (0.007) | (0.007) |
| Strength of financial regulations | -0.045*** | 0.010 | 0.010 |
| | (0.014) | (0.011) | (0.011) |
| Enabling financial environment | 0.036*** | 0.003 | 0.002 |
| | (0.012) | (0.005) | (0.005) |
| Base sci & tech | -0.179*** | -0.023 | -0.028 |
| | (0.034) | (0.019) | (0.018) |
| Medium sci & tech | -0.127*** | -0.001 | 0.002 |
| | (0.017) | (0.009) | (0.008) |
| High sci & tech | -0.061*** | -0.002 | 0.001 |
| | (0.014) | (0.006) | (0.006) |
| Social capacity (incl. equity, inclusion, etc.) | -0.128*** | 0.004 | -0.001 |
| | (0.022) | (0.013) | (0.012) |
| Constant | 7.329*** | 7.181*** | 7.149*** |
| | (0.045) | (0.045) | (0.020) |
| | | | |
| Observations | 1,230 | 1,230 | 1,230 |
| R-squared | 0.799 | 0.727 | 0.468 |
| Controls | YES | YES | YES |
| Year Fixed Effects | YES | YES | YES |
| Country Fixed Effects | NO | NO | YES |
| Number of countries | 82 | 82 | 82 |

Standard errors in parentheses
*** p<0.01, ** p<0.05, * p<0.1

**Appendix J. Table. Sensitivity Analysis, Only Fixed Effects Regressions—With/Without Time Effects, With/Without Controls, Robust/Standard Errors. Dependent Variable, Log of GDP Per Capita.**

| VARIABLES | Model 1 | Model 2 | Model 3 | Model 4 |
|---|---|---|---|---|
| Public policy (inc. fiscal, monetary, structural…) | 0.067*** | 0.070*** | 0.087*** | 0.087*** |
| | (0.012) | (0.013) | (0.013) | (0.027) |
| Infrastructure (ICT & energy) | 0.161*** | 0.164*** | 0.095*** | 0.095*** |
| | (0.012) | (0.013) | (0.017) | (0.028) |
| Logistic Per. Index (trade & transp. infras.) | 0.054*** | 0.053*** | 0.029*** | 0.029*** |
| | (0.006) | (0.006) | (0.007) | (0.009) |
| Specialized skills | 0.089*** | 0.092*** | 0.061*** | 0.061*** |
| | (0.016) | (0.015) | (0.016) | (0.018) |
| Generalized skills | -0.027*** | -0.027*** | -0.018** | -0.018 |
| | (0.009) | (0.009) | (0.009) | (0.014) |
| Financial infrastructure | 0.030*** | 0.028** | 0.025** | 0.025** |
| | (0.011) | (0.011) | (0.011) | (0.012) |
| Financial environment | 0.004 | 0.009 | 0.025*** | 0.025** |
| | (0.007) | (0.007) | (0.007) | (0.010) |
| Strength of financial regulations | 0.032*** | 0.019* | 0.010 | 0.010 |
| | (0.011) | (0.011) | (0.011) | (0.018) |
| Enabling financial environment | 0.001 | 0.001 | 0.002 | 0.002 |
| | (0.005) | (0.005) | (0.005) | (0.005) |
| Base sci & tech | 0.037*** | -0.029 | -0.028 | -0.028 |
| | (0.009) | (0.018) | (0.018) | (0.025) |
| Medium sci & tech | -0.009 | -0.007 | 0.002 | 0.002 |
| | (0.008) | (0.008) | (0.008) | (0.013) |
| High sci & tech | -0.003 | -0.002 | 0.001 | 0.001 |
| | (0.006) | (0.006) | (0.006) | (0.006) |
| Social capacity (incl. equity, inclusion, etc.) | 0.019 | 0.007 | -0.001 | -0.001 |
| | (0.012) | (0.012) | (0.012) | (0.018) |
| Tech. coop. grants (st) | | -0.022** | -0.017 | -0.017 |
| | | (0.011) | (0.011) | (0.018) |
| Total population (st) | | 0.646*** | 0.388** | 0.388 |
| | | (0.176) | (0.180) | (0.345) |
| Gross capital (st) | | -0.014 | 0.035 | 0.035 |
| | | (0.044) | (0.045) | (0.063) |
| Incoming tourists' no. (st) | | 0.017* | 0.021** | 0.021 |
| | | (0.010) | (0.010) | (0.013) |
| Merch. imports frm HICs (st) | | 0.046*** | 0.058*** | 0.058*** |
| | | (0.012) | (0.012) | (0.022) |
| Net ODA/aid received (st) | | 0.004 | -0.001 | -0.001 |
| | | (0.009) | (0.009) | (0.019) |
| Health expenditure (st) | | -0.052*** | - | - |

| VARIABLES | Model 1 | Model 2 | Model 3 | Model 4 |
|---|---|---|---|---|
| | | | 0.050*** | 0.050*** |
| | | (0.009) | (0.009) | (0.016) |
| No. of employers (st) | | 0.023** | 0.029*** | 0.029*** |
| | | (0.009) | (0.009) | (0.010) |
| YR2006 | | | 0.011 | 0.011 |
| | | | (0.021) | (0.014) |
| YR2007 | | | 0.033 | 0.033* |
| | | | (0.021) | (0.017) |
| YR2008 | | | 0.053** | 0.053*** |
| | | | (0.022) | (0.017) |
| YR2009 | | | 0.040* | 0.040* |
| | | | (0.022) | (0.022) |
| YR2010 | | | 0.071*** | 0.071*** |
| | | | (0.023) | (0.023) |
| YR2011 | | | 0.080*** | 0.080*** |
| | | | (0.024) | (0.026) |
| YR2012 | | | 0.100*** | 0.100*** |
| | | | (0.025) | (0.034) |
| YR2013 | | | 0.132*** | 0.132*** |
| | | | (0.026) | (0.041) |
| YR2014 | | | 0.134*** | 0.134*** |
| | | | (0.027) | (0.042) |
| YR2015 | | | 0.128*** | 0.128** |
| | | | (0.028) | (0.050) |
| YR2016 | | | 0.135*** | 0.135** |
| | | | (0.029) | (0.052) |
| YR2017 | | | 0.155*** | 0.155*** |
| | | | (0.031) | (0.056) |
| YR2018 | | | 0.147*** | 0.147** |
| | | | (0.032) | (0.057) |
| YR2019 | | | 0.166*** | 0.166** |
| | | | (0.032) | (0.065) |
| Constant | 7.241*** | 7.241*** | 7.149*** | 7.149*** |
| | (0.004) | (0.004) | (0.020) | (0.032) |
| Observations | 1,230 | 1,230 | 1,230 | 1,230 |
| R-squared | 0.407 | 0.448 | 0.468 | 0.468 |
| Number of countryname1 | 82 | 82 | 82 | 82 |
| Controls | NO | YES | YES | YES |
| Country Fixed Effects | YES | YES | YES | YES |
| Year Fixed Effects | NO | NO | YES | YES |
| Errors | SE | SE | SE | Robust |

Standard errors in parentheses
*** p<0.01, ** p<0.05, * p<0.1

**Appendix K. Table. Sensitivity Analysis: Data Averaged over 5 Years Period. Dependent Variable, Log of GDP Per Capita.**

| VARIABLES | Pooled OLS | Random Effects | Fixed Effects |
|---|---|---|---|
| Public policy (inc. fiscal, monetary, structural…) | 0.100*** | 0.071*** | 0.080*** |
| | (0.022) | (0.024) | (0.026) |
| Infrastructure (ICT & energy) | 0.363*** | 0.150*** | 0.122*** |
| | (0.025) | (0.019) | (0.021) |
| Logistic Per. Index (trade & transp. infras.) | 0.095*** | 0.043*** | 0.038*** |
| | (0.016) | (0.007) | (0.007) |
| Specialized skills | 0.242*** | 0.120*** | 0.076*** |
| | (0.024) | (0.018) | (0.018) |
| Generalized skills | -0.083*** | -0.032** | -0.022 |
| | (0.012) | (0.015) | (0.014) |
| Financial infrastructure | 0.109*** | 0.038*** | 0.027** |
| | (0.017) | (0.013) | (0.012) |
| Financial environment | 0.050*** | 0.018* | 0.018* |
| | (0.015) | (0.010) | (0.010) |
| Strength of financial regulations | -0.045*** | 0.014 | 0.016 |
| | (0.014) | (0.018) | (0.018) |
| Enabling financial environment | 0.035*** | 0.003 | 0.001 |
| | (0.011) | (0.005) | (0.005) |
| Base sci & tech | -0.179*** | -0.020 | -0.028 |
| | (0.034) | (0.024) | (0.026) |
| Medium sci & tech | -0.124*** | -0.003 | -0.001 |
| | (0.017) | (0.012) | (0.012) |
| High sci & tech | -0.061*** | -0.004 | -0.001 |
| | (0.014) | (0.006) | (0.005) |
| Social capacity (incl. equity, inclusion, etc.) | -0.127*** | 0.006 | 0.002 |
| | (0.023) | (0.019) | (0.018) |
| period = 2010-15 | -0.049 | 0.040** | 0.054*** |
| | (0.030) | (0.019) | (0.018) |
| period = 2015-19 | -0.096*** | 0.056* | 0.077** |
| | (0.036) | (0.031) | (0.033) |
| Constant | 7.289*** | 7.209*** | 7.198*** |
| | (0.024) | (0.061) | (0.016) |
| Observations | 1,230 | 1,230 | 1,230 |
| R-squared | 0.799 | 0.729 | 0.458 |
| Controls | YES | YES | YES |
| Year Fixed Effects | YES | YES | YES |
| Country Fixed Effects | NO | NO | YES |
| Robust Standard Errors | YES | YES | YES |
| Number of countries | 82 | 82 | 82 |

Standard errors in parentheses
*** p<0.01, ** p<0.05, * p<0.1